\documentclass[useAMS,usenatbib]{mn2e} 
\usepackage{epsfig}  
\usepackage{textcomp} 

\def\mums{\textmu m }
\def\mum{\textmu m}
\def\isos{{\it ISO }}
\def\iso{{\it ISO}}

\title[A study of the 15 \mums quasars in the ELAIS {\it N1} and {\it N2} fields]
{A study of the 15 \mums quasars in the ELAIS {\it N1} and {\it N2} fields}  
       
\author[Afonso-Luis et al.]{A. Afonso-Luis$^{1}$, E. Hatziminaoglou$^{1}$, 
I. P\'{e}rez-Fournon$^{1}$, E.A. Gonz\'{a}lez-Solares$^{2}$,
\newauthor M. Rowan-Robinson$^{3}$, M. Vaccari$^{3}$, C. Lari$^{4}$,  
S. Serjeant$^{5}$, S. Oliver$^{6}$, 
\newauthor A. Hern\'{a}n-Caballero$^{1}$, F.M. Montenegro-Montes$^{1}$\\  
$^{1}$Instituto de Astrofisica de Canarias, C/ Via Lactea s/n, E-38200 La Laguna, Spain\\      
$^{2}$Institute of Astronomy, University of Cambridge, Madingley Road, Cambridge CB3 0HA, UK\\
$^{3}$Imperial College of Science, Technology \& Medicine, Prince Consort Road, London SW7 2BZ, UK\\
$^{4}$Instituto de Radioastronomia, Via P. Gobetti 101, Bologna 40129, Italy\\
$^{5}$Centre for Astrophysics and Planetary Science, School of Physical Sciences, University of Kent, 
Canterbury, Kent CT2 7NR, UK\\
$^{6}$Astronomy Centre, Department of Physics and Astronomy, University of Sussex, Falmer, 
Brighton BN1 9QJ, UK}        
  
\begin{document}       
  
\maketitle  
  
\begin{abstract}   
\noindent
This paper discusses properties of the European Large Area \isos Survey 
15 \mums quasars and tries to establish
a robust method of quasar selection for future use within the Spitzer Wide-Area
Infrared Extragalactic Survey (SWIRE) framework. The importance
of good quality ground-based optical data is stressed, both for the candidates selection and for
the photometric redshifts estimates. Colour-colour plots and template fitting are used for
these purposes. The properties of the 15 \mums quasars sample are studied, including 
variability and black hole masses and compared
to the properties of other quasars that lie within the same fields but have no mid-infrared
counterparts. The two subsamples do not present substantial differences and are
believed to come from the same parent population.
\end{abstract}  
  
\maketitle  
     
\section{INTRODUCTION}
\label{elais}         

The European Large Area \isos Survey (ELAIS) 
\citep{oliver00} is the largest survey performed 
with the Infrared Space Observatory \citep{kessler96} at 6.7, 15, 90 and 175 \mums
and resulted in the delivery of the largest catalogue of any \isos survey
\citep{rowan03} from both the ISOCAM \citep{cesarsky96} and ISOPHOT
\citep{lemke96} instruments.
In particular, the 15 \mums survey (performed with the ISOCAM instrument) 
covers an area of $\sim$12 deg$^2$, divided into four main fields ({\it N1}, 
{\it N2}, {\it N3}, and {\it S1}) and several smaller areas. 15 \mums
observations in the four main fields were analysed by \cite{vaccari04},
providing a catalogue of 1923 sources detected with $S/N > 5$ over 10.85 deg$^2$.
The Final Band-Merged Catalogue \citep{rowan03} combines all source lists
at different wavelengths and redshifts obtained to date in ELAIS fields. This catalogue 
comprises a total of 3523 entries with about one
third having spectroscopic identifications.

Due to the fact that complete spectroscopic follow-up is usually not feasible over large 
and deep fields, one needs to use tools for detecting quasar candidates using 
photometric data only. As part of our study of mid-infrared (IR) quasars we present 
the results of two independent quasar candidates
selection techniques, one based on colour-colour diagrams, and the other one
template fitting, and try to improve the selection including IR constraints.
In recent years, with the available multicolour surveys, quasar photometric 
redshift methods have been developed, yielding however somewhat
less reliable results than the ones for galaxies (e.g. \citealt{hatzi00};
\citealt{richards01}). 
For the purposes of this work, the template fitting technique \citep{hatzi00} 
is applied on the two different
data sets available for the studied fields (SDSS and WFS).
All methods and results described in this paper can be directly applied
to the Spitzer Wide-Area Infrared Extragalactic Survey (SWIRE; \citealt{lonsdale03}).

The layout of this paper is as follows. In Section \ref{knownqsos} a brief description of 
the optical data available for this work is given. Section \ref{variability}
deals with variability issues. In Section \ref{candidates} a description of the selection 
of quasar candidates using optical and IR properties is given, based
on two different methods. Photometric redshifts for the spectroscopically confirmed
quasars in the two data sets (SDSS and WFS) are estimated and the results obtained for 
the two photometric systems are discussed. Section \ref{comparison} compares the
results for the sources with and without IR counterparts, in terms of their statistical
properties and black hole (BH) masses. Our conclusions are presented in Section \ref{discuss}.

\section{THE 15 \mums QUASAR SAMPLE AND RELATED OPTICAL DATA}
\label{knownqsos}

In the present work we study the type I quasars detected by \isos at 15 \mums in two of 
the ELAIS fields, {\it N1} and {\it N2}. Throughout this work, the traditional (but
conservative) requirement of a quasar to be point-like has been set and sources classified 
as extended based on their $r$-band morphology are not taken into account.
The morphological selection is made in order to avoid contamination by galaxies of the 
quasar candidates samples discussed in Section \ref{candidates} and has no impact on the
validity of the results presented here.
Out of the 1056 sources in {\it N1} and {\it N2} contained in the ELAIS 15 $\mu$m Final 
Analysis Catalogue Version 1.0 \citep{vaccari04}, 849 sources were 
identified in INT WFS images by \cite{gonzalez04}, the 
non-identification being due either to incomplete optical coverage of the 
15 \mums fields or the optical limiting magnitude.

The ELAIS {\it N1} and {\it N2} have been fully 
covered by the Wide Field Survey (WFS; \cite{mcmahon01}, carried out with the prime focus 
Wide Field Camera (WFC) at the 2.5m Isaac Newton Telescope (INT) at La Palma. 
The survey consists of single 600s exposures in five bands {\it U}, {\it g},
{\ r}, {\it i} and {\it z} down to AB magnitude limits (5$\sigma$ limits for a 
point source) of 24.1, 24.8, 24.1, 23.6
and 22.4, respectively. The AB corrections for the conversion from the (original) Vega 
magnitudes to AB magnitudes have been computed using HyperZ \citep{micol00} and are:
0.751, -0.063, 0.165, 0.407 and 0.534, for {\it U}, {\it g},
{\ r}, {\it i} and {\it z}, respectively.
Out of the 849 sources detected in 15 \mums in {\it N1} and {\it N2}, there are 110
point sources (SExtractor CLASS\_STAR $\ge 0.9$),
excluding saturated stars, with emission at 15 \mums \citep{gonzalez04}.
    
In addition to WFS, the Sloan Digital Sky Survey has validated and made publicly
available its Data Release 1 (DR1; \citealt{abazajian03}), partially covering
our fields. SDSS consists
of five-band ($u$, $g$, $r$, $i$, $z$) imaging data covering 2099 deg$^2$, 186.240 spectra of
galaxies, quasars, stars and calibrating blank sky patches selected
over 1360 deg$^2$ of this area, and catalogues of measured parameters from
these data. The imaging data reach a limiting AB magnitude \citep{vanden04}
of $r \sim 22.6$ (95\% completeness limit for stars). Among the 15 \mums sources
in the areas covered by the SDSS DR1 photometry, 82 have been
morphologically classified as point sources (SDSS OBJC\_TYPE = 6).
In fact, the partial coverage of {\it N1} and {\it N2} provides photometry and spectra
for a variety of sources. More particularly,
among the 36 SDSS spectroscopically confirmed quasars lying in the {\it N1} and {\it N2} fields,
16 have been detected at 15 \mum. Note, however, that the areas covered by spectroscopy
do not exactly coincide with those covered by photometry and are, in fact, smaller.
Two morphologically extended low-redshift quasars (0.214 and 0.245) with $r$-band magnitudes 
of 18.25 and 18.58, have been excluded from the spectroscopically confirmed quasar sample, for the
reasons mentioned at the beginning of this Section.
Throughout this work and unless otherwise stated, all magnitudes will refer to the AB system.

\begin{figure*}
\centerline{
\psfig{file=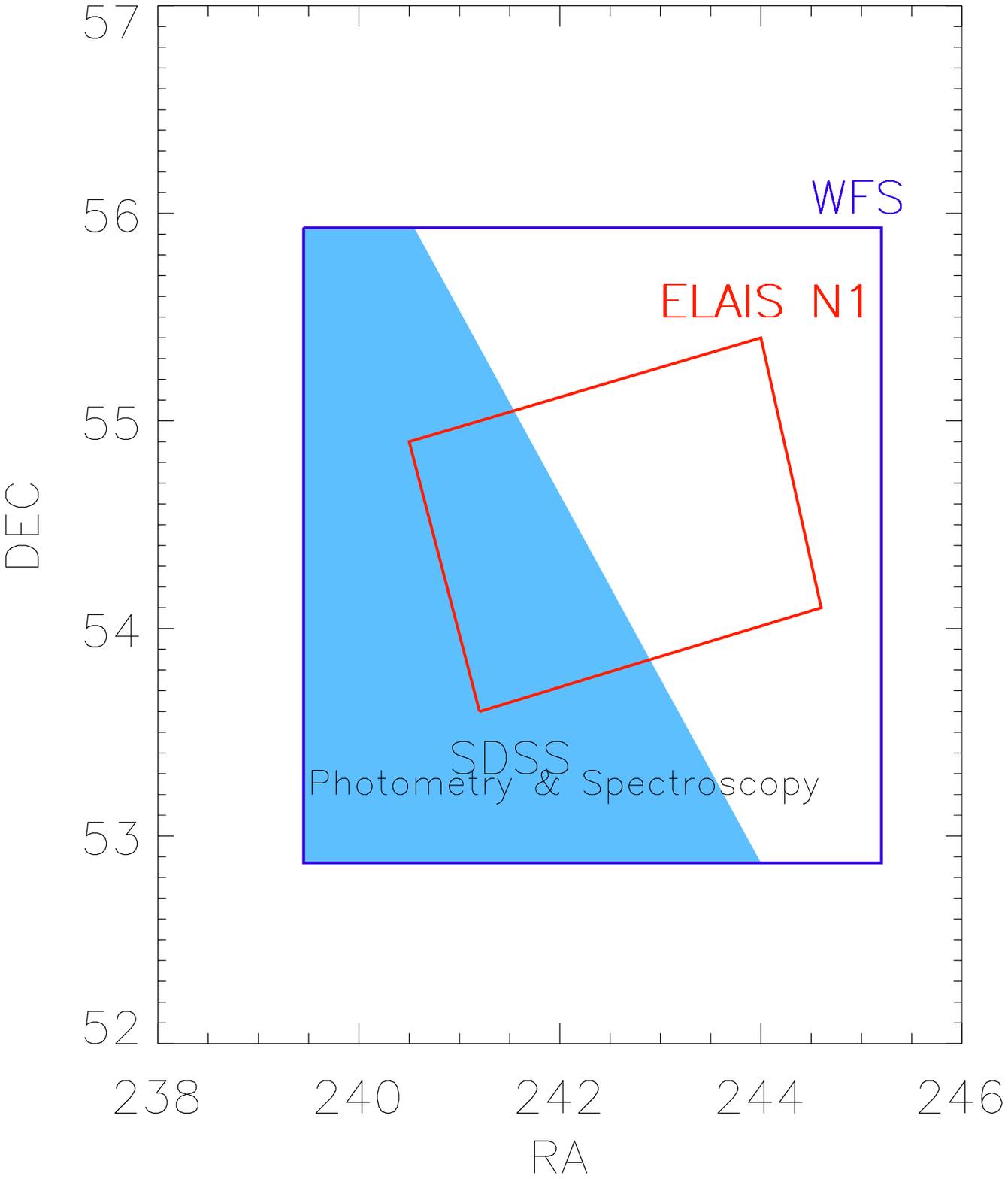,width=7cm,height=6.5cm}
\psfig{file=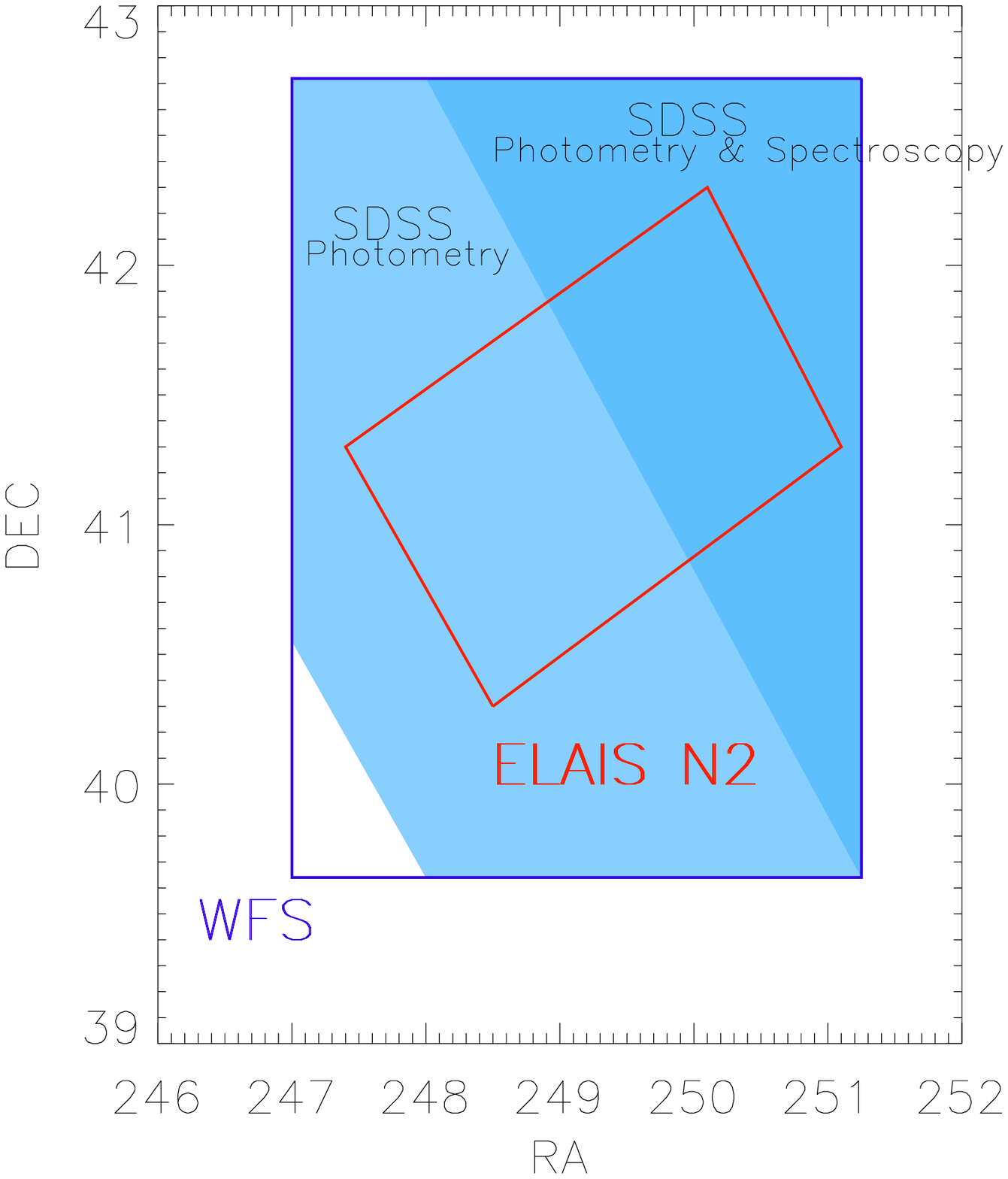,width=7cm,height=6.5cm}}
\caption{Coverage of the ELAIS {\it N1} and {\it N2} fields (small inner rectangles) 
by the WFS and the SDSS DR1 photometric and spectroscopic data.}
\label{figcoverage}
\end{figure*}

Fig. \ref{figcoverage} shows the coverage of the ELAIS {\it N1} and {\it N2} fields
by the WFS and the SDSS DR1 photometric and spectroscopic data. 

Finally, as a part of the extensive program of ground based spectroscopy associated 
with the ELAIS survey, there have been follow-up observations of ELAIS 15 \mums sources in the 
{\it N1} and {\it N2} fields (P\'{e}rez-Fournon et al. in preparation) using the WYFFOS multi-fibre spectrograph, on
the William Herschel Telescope (WHT) at La Palma. This observing run added 
nine previously unknown quasars to the spectroscopic sample 
in the regions of {\it N1} and {\it N2} not covered by the spectroscopic DR1.

\begin{figure}
\centerline{
\psfig{file=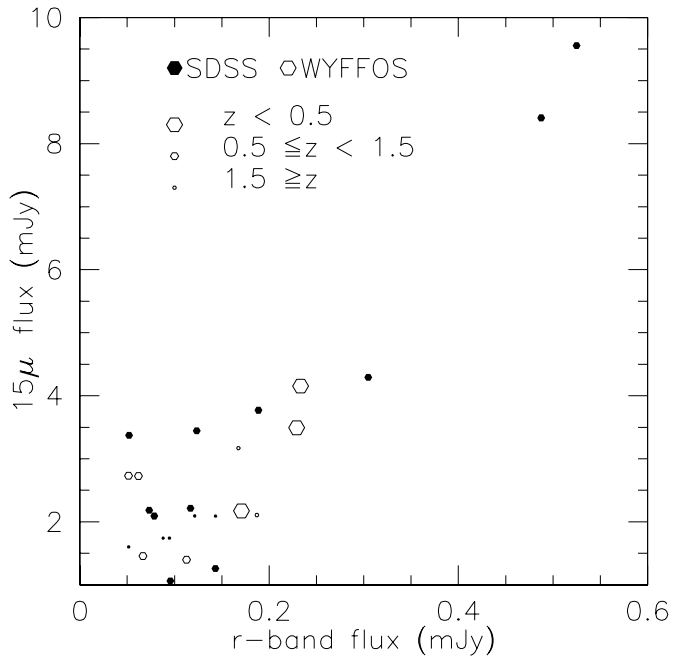,width=9cm,height=9cm}}
\caption{$r$-band versus the 15 \mums flux for the quasar sample. Filled (open) circles represent
objects with SDSS (WYFFOS) spectra. Symbol size decreases with increasing redshift.}
\label{figflux}
\end{figure}

Therefore, a total of 25 spectroscopically confirmed quasars have been identified
among the final ELAIS 15 \mums catalogue and their properties are described in
Table \ref{tabconfquasars}. 
Fig. \ref{figflux} shows the 15 \mums flux versus $r$-band flux for the quasar sample.

\begin{table*}
\caption{ELAIS 15 \mums spectroscopically confirmed quasars. $r$-band magnitude
in column 5 is the SDSS $r$-band magnitude apart from the objects marked with
an asterisk. These objects are outside the spectroscopic coverage of DR1 and the
magnitude shown is the 
WFS $r$ magnitude. The last column (reference) refers to the origin of the
spectra (1: \citealt{abazajian03}; 2: P\'{e}rez-Fournon et al. in preparation)}
\label{tabconfquasars}
\begin{tabular}{ccccccc}
\hline
\hline
ELAIS ID&RA (optical)&Dec (optical)&z&$r$ mag& 15 \mums \, flux (mJy)&$z$ Reference\\
\hline
ELAISC15\_J160250.9+545057&240.71234131&54.84947205&1.1971&19.22&2.1820&SDSS$^1$\\
ELAISC15\_J160522.9+545613&241.34640503&54.93708420&0.5722&18.94&2.2160&SDSS$^1$\\
ELAISC15\_J160623.5+540556&241.59822083&54.09888840&0.8766&17.62&4.2930&SDSS$^1$\\
ELAISC15\_J160630.5+542007&241.62754822&54.33544159&0.8205&18.73&3.4460&SDSS$^1$\\
ELAISC15\_J160638.0+535009&241.65779114&53.83572006&2.9426&19.77&1.6010&SDSS$^1$\\
ELAISC15\_J161007.2+535814&242.52960205&53.97055435&2.0317&18.86&1.7400&SDSS$^1$\\
ELAISC15\_J163702.2+413022&249.25930786&41.50616837&1.1783&19.11&2.0920&SDSS$^1$\\
ELAISC15\_J163709.2+414031&249.28890991&41.67519760&0.7602&17.20&8.4090&SDSS$^1$\\
ELAISC15\_J163739.3+414348&249.41436768&41.72999954&1.4136&18.94&1.0610&SDSS$^1$\\
ELAISC15\_J163847.5+421141&249.69760132&42.19494629&1.7786&18.93&1.7420&SDSS$^1$\\
ELAISC15\_J163915.9+412834&249.81591797&41.47602844&0.6919&19.05&3.3730&SDSS$^1$\\
ELAISC15\_J163930.8+410013&249.87844849&41.00380707&1.0515&18.23&1.2580&SDSS$^1$\\
ELAISC15\_J163952.9+410346&249.97023010&41.06244278&1.6050&18.58&2.0940&SDSS$^1$\\
ELAISC15\_J164010.1+410521&250.04225159&41.08955383&1.0990&17.01&9.5570&SDSS$^1$\\
ELAISC15\_J164016.0+412102&250.06701660&41.35038757&1.7570&18.44&2.0900&SDSS$^1$\\
ELAISC15\_J164018.4+405812&250.07641602&40.97030640&1.3175&18.13&3.7700&SDSS$^1$\\
ELAISC15\_J161521.8+543148&243.84077454&54.53016663&0.4737&18.24*&3.492&WYFFOS$^2$\\
ELAISC15\_J161526.7+543004&243.86094666&54.50175095&1.3670&19.35*&1.457&WYFFOS$^2$\\
ELAISC15\_J161543.5+544828&243.93133545&54.80799866&1.6920&18.22*&2.107&WYFFOS$^2$\\
ELAISC15\_J163425.2+404152&248.60472107&40.69794464&1.6840&18.37&3.1690&WYFFOS$^2$\\
ELAISC15\_J163502.7+412953&248.76176453&41.49808502&0.4727&18.08&2.1730&WYFFOS$^2$\\
ELAISC15\_J163531.1+410025&248.87956238&41.00761032&1.1500&18.81&1.3990&WYFFOS$^2$\\
ELAISC15\_J163533.9+404025&248.89176941&40.67380524&0.5340&19.72&2.7300&WYFFOS$^2$\\
ELAISC15\_J163553.5+412054&248.97354126&41.34883118&1.1950&19.39&2.7260&WYFFOS$^2$\\
ELAISC15\_J163634.4+412742&249.14343262&41.46200180&0.1711&18.50&4.1540&WYFFOS$^2$\\
\hline
\hline
\end{tabular}
\end{table*}

\section{Variability}         
\label{variability}

Variations in the luminosity of quasars have been observed from X-ray to radio wavelengths, 
with timescales of minutes to years. The majority of QSOs have 
continuum variability on the order of 10\% on timescales of months to years \citep{vanden04}.
Furthermore, recent observations of radio-quiet quasars indicate that more than 80\% 
show long-term (month to year) variability with amplitudes up to half a magnitude 
\citep{huber02}. Variability is wavelength dependent. The continuum 
($F_{\nu} \propto \nu^{-\alpha}$) tends to get harder 
(the spectral index, $\alpha$, decreases) as the quasar gets brighter, which means that the variations
are larger at shorter wavelengths.

\begin{figure*}
\centerline{
\psfig{file=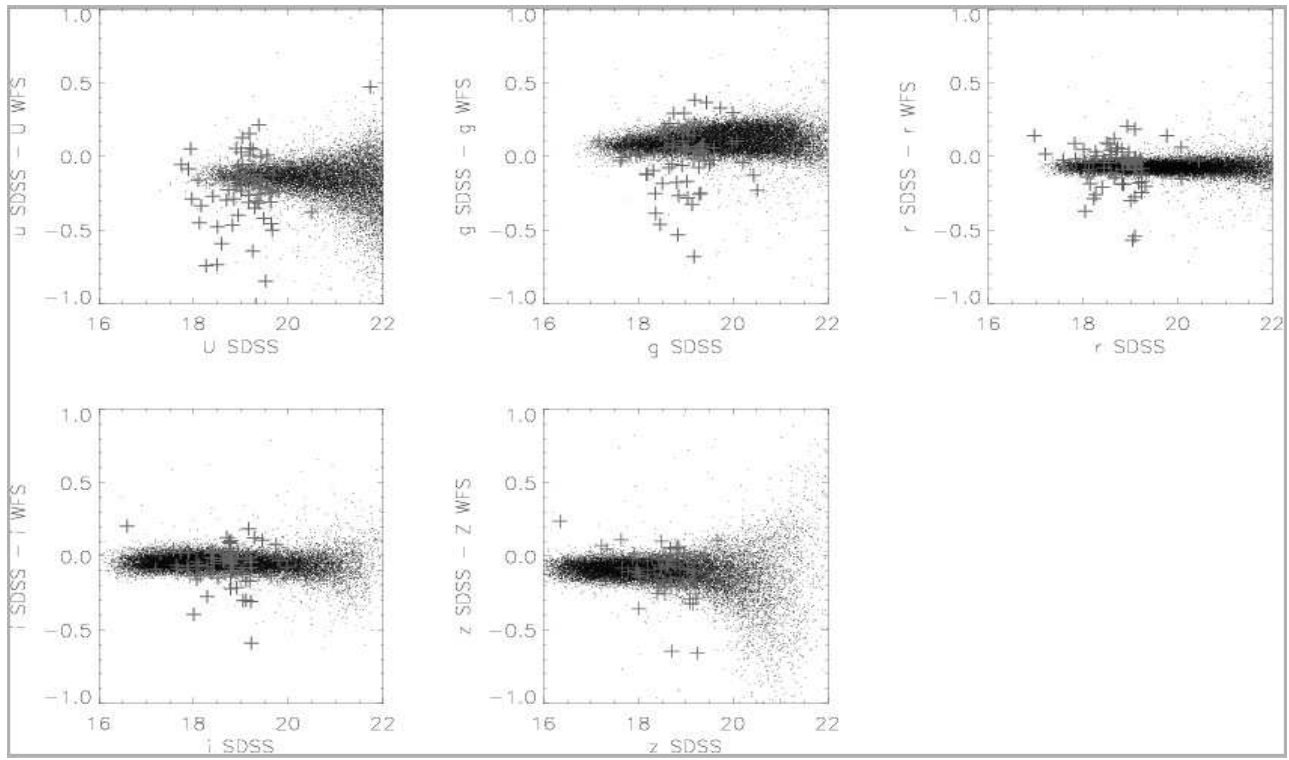,width=17cm,height=10cm}\qquad}
\caption{Magnitudes differences between SDSS and WFS, the black points 
represent the point sources, mostly stars, while the red crosses indicate
the 73 confirmed quasars. The photometric errors for the quasars are lower than 
0.05 magnitudes in all bands and the dispersion in their magnitudes is evident.}
\label{figdifmag}
\end{figure*}

The SDSS imaging strategy consists in observing in almost simultaneous mode
the same field in the five different bands. The imaging strategy of the WFS 
was, however, different, and the same fields 
were observed in the different filters in timescales ranging from a few months
to more than a year. Fig. \ref{figdifmag} shows the differences of the magnitudes 
for the point sources lying in the common area in the two photometric 
systems. The magnitude differences show very small dispersions down to a certain magnitude 
for almost all point sources that are identified as stars
but large variations appear in the case of many of the confirmed quasars
(red crosses) in all filters but especially in the $U$- and $g$-bands, indicating variability
related issues.

\begin{figure}
\centerline{
\psfig{file=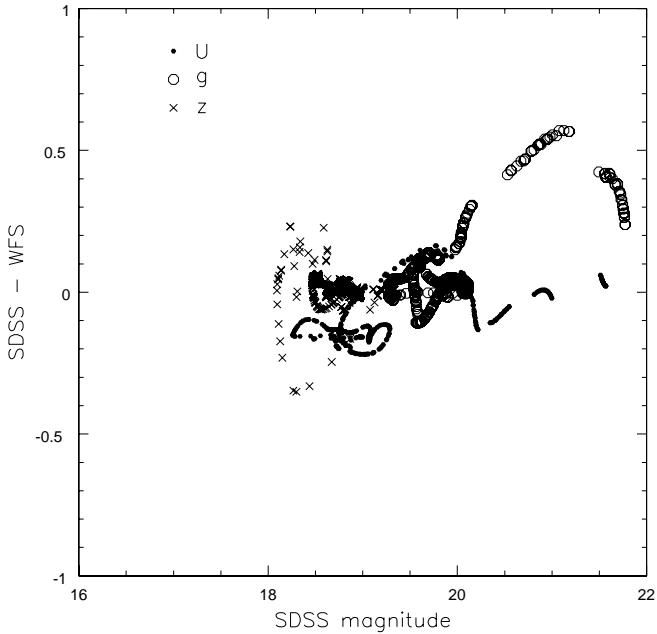,width=9cm,height=9cm}}
\caption{Theoretical magnitude differences for 
the filters {\it U}, {\it g}, and {\it z} of the SDSS and WFS photometric
systems, in the redshift interval [0,4].
Filters $r$ and $i$ have been omitted for clarity.}
\label{figtheordifmag}
\end{figure}

It might be argued that the variations are due to the different photometric systems, however close
they may be. For this purpose, a simulated quasar catalogue was created, based on the colours of the SDSS 
composite quasar spectrum \citep{vanden01}, including all 10 filters and for redshifts
spanning from 0 to 6 and the magnitudes were compared. Fig. \ref{figtheordifmag} illustrates the
expected magnitude differences as a function of magnitude for redshifts in the interval [0,4], 
where all our spectroscopically confirmed quasars lie, and for the filters $U$, $g$ and $z$. 
Filters $r$ and $i$ were omitted for clarity as the points largely overlap with the rest of
the points. Even though magnitude variations are predicted, their amplitude is
significantly smaller than the ones observed. We therefore conclude that not all 
observed variations between the SDSS and WFS photometry can be explained by the differences of the 
photometric systems and that some must be a result of quasar variability.

This strongly suggests that quasar candidates selection and photometric redshift estimates
are likely not to be accurate for all objects when WFS or other similar data are used
and additional methods have to be looked for.

\section{QUASAR CANDIDATES SELECTION}
\label{candidates}   

For the purpose of identifying high probability quasar candidates from the available datasets
we used two techniques,
a combination of colour-colour diagrams (hereafter method 1 - {\it M1}) and the template fitting
method ({\it M2}).
{\it M1}, based on \cite{richards02}, consists of a colour-colour selection algorithm 
trained using the SDSS Early Data Release \citep{stoughton02}, for low- to
intermediate redshift ($\sim 2.5$) objects. {\it M2} is the standard 
template fitting that simultaneously provides a photometric redshift estimate for
the quasar candidates.
The point source template library comprises quasar and stellar templates and the
observed SED of each object is compared to the one computed by convolving each
template with the filters transmissions (for further details see
\citealt{hatzi00} and 2002).
{\it M1} is expected to give higher confirmation rates (i.e. number of real quasars
over then number of quasar candidates) at low and intermediate redshifts
but its efficiency greatly depends on the photometric system used and
can not be applied as such when other filters are used.
{\it M2} is subject to higher contamination from sources other than quasars (e.g. stars)
but is expected to have a much better efficiency at high redshift and can be used
for any filter combination. In order to improve the results, we also make use of the IR information
available for the 15 \mums sources keeping in mind that the combination of all these techniques
will be applied in the near future in the framework of SWIRE.

A first test is made using the more reliable SDSS photometry (Section \ref{sdsscand}) but
an attempt of selecting candidates based on the WFS photometry will also be presented
(Section \ref{wfscand}). Our test sample consists of 21 spectroscopically confirmed
quasars with available SDSS photometry and 25 with WFS photometry,
which include all 21 from SDSS (see Table \ref{tabconfquasars}).
Fine-tuning the method for WFS data is very important as large part of
the SWIRE fields have been observed by the WFS.

\subsection{Quasar Candidates with SDSS Photometry}   
\label{sdsscand}

A sample of 82 15 \mums ELAIS sources identified as point sources from their optical photometry
(SDSS OBJC\_TYPE = 6) with $r$-band magnitudes brighter that 22.6
has been selected from the SDSS photometric catalogue. The optical magnitude cut has been
imposed in order to avoid spurious detections and large photometric errors.

\begin{table}
\caption{Comparison of the completeness and confirmation rate yielded by {\it M1} and {\it M2} on
a sample of 82 SDSS point sources with 15 \mums emission. The second column shows the number of
quasar candidates ($N_c$) and the number of spectroscopically confirmed quasars among them ($N_f$).}
\label{tabcand}
\begin{tabular}{lccc}
\hline
\hline
Method & $N_c$ ($N_f$) & Completeness & Conf. Rate \\
\hline
\hline
{\it M1}     & 26 (19) & 90\%         & 73\%       \\
{\it M2}     & 33 (19) & 90\%         & 58\%       \\
\hline
{\it M2} \& {\it C1} & 30 (19) & 90\% & 63\%     \\
\hline
\hline
\end{tabular}
\end{table}

\begin{figure*}
\centerline{
\psfig{file=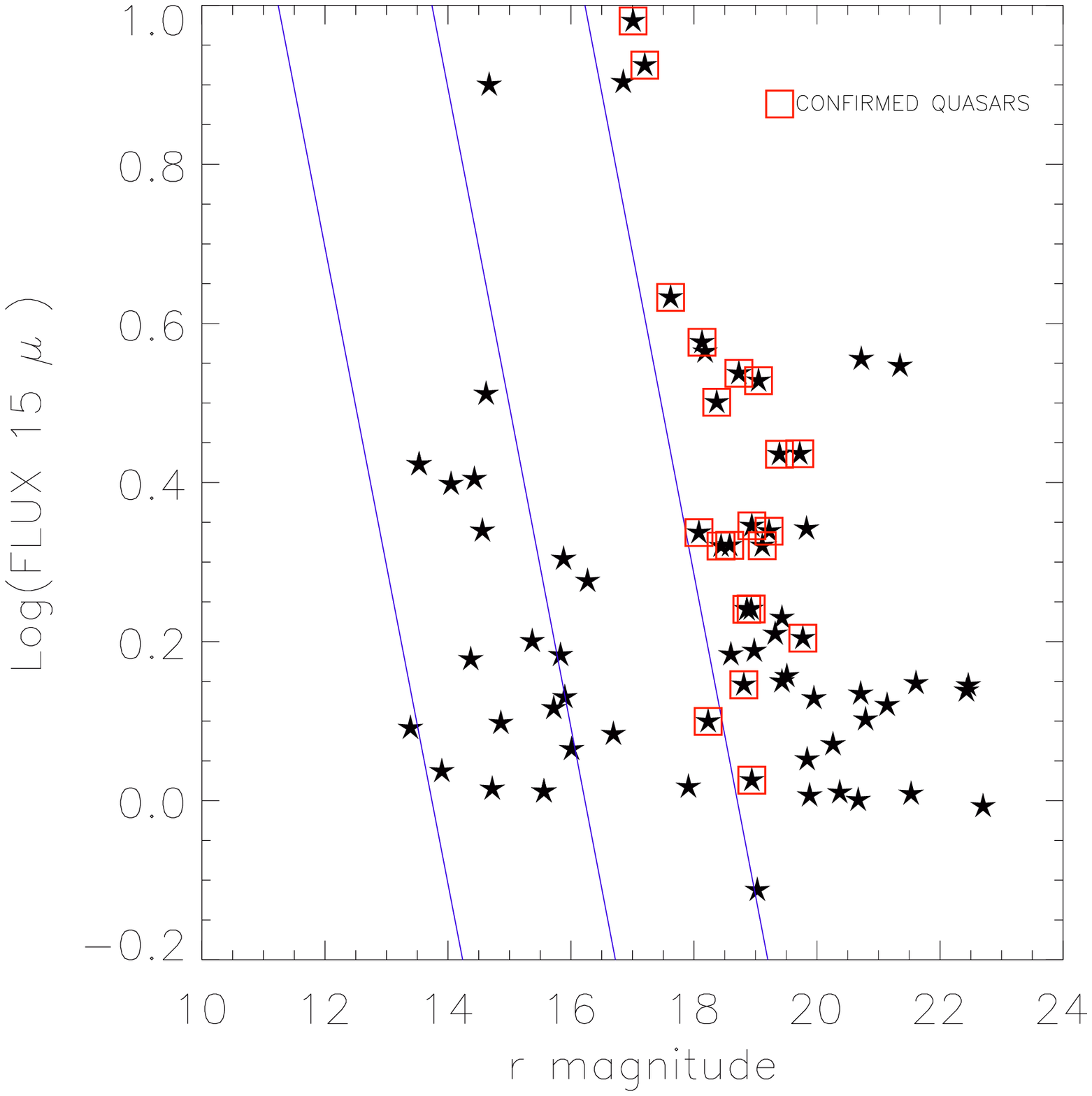,width=6cm,height=6cm}
\psfig{file=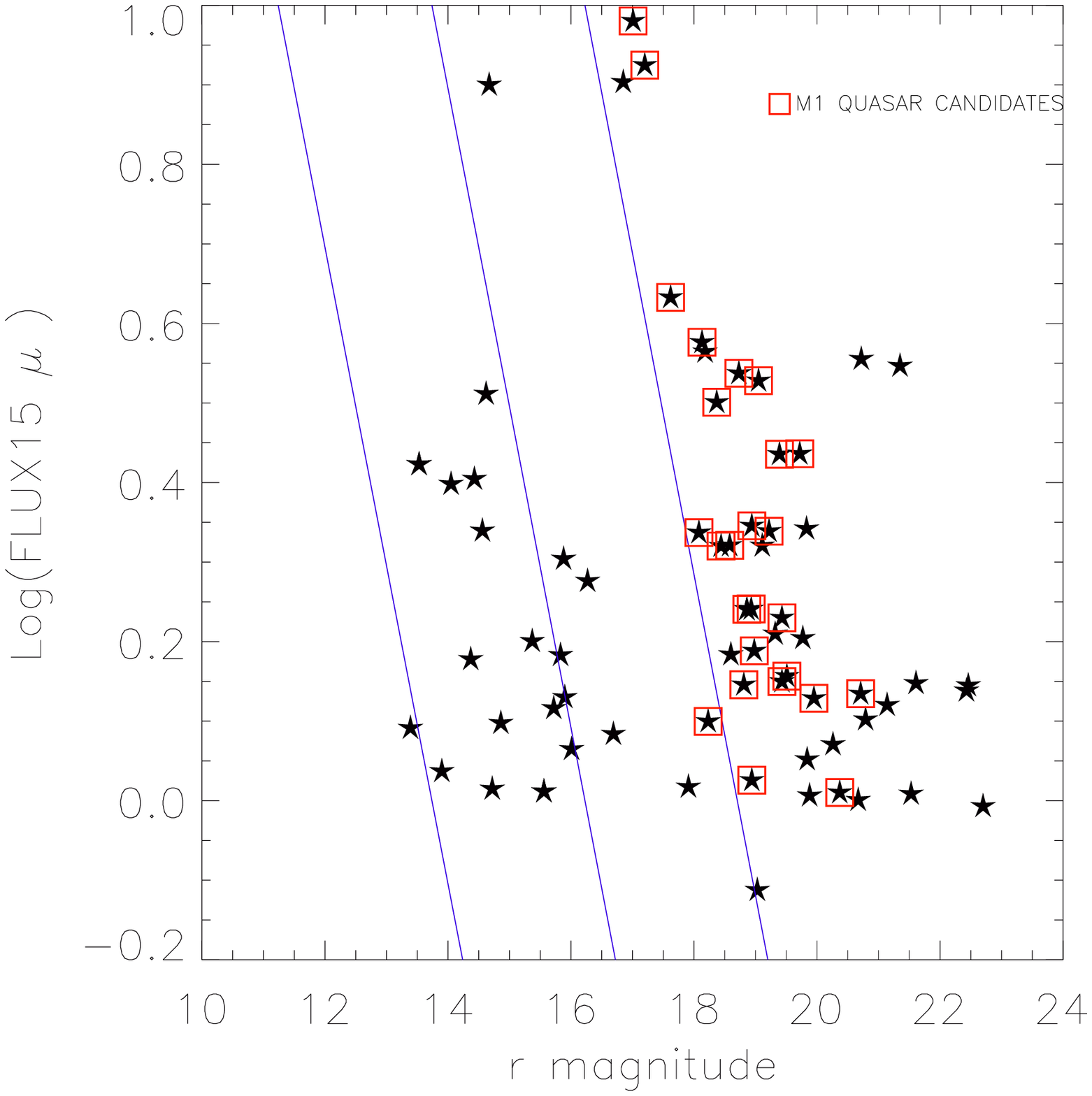,width=6cm,height=6cm}
\psfig{file=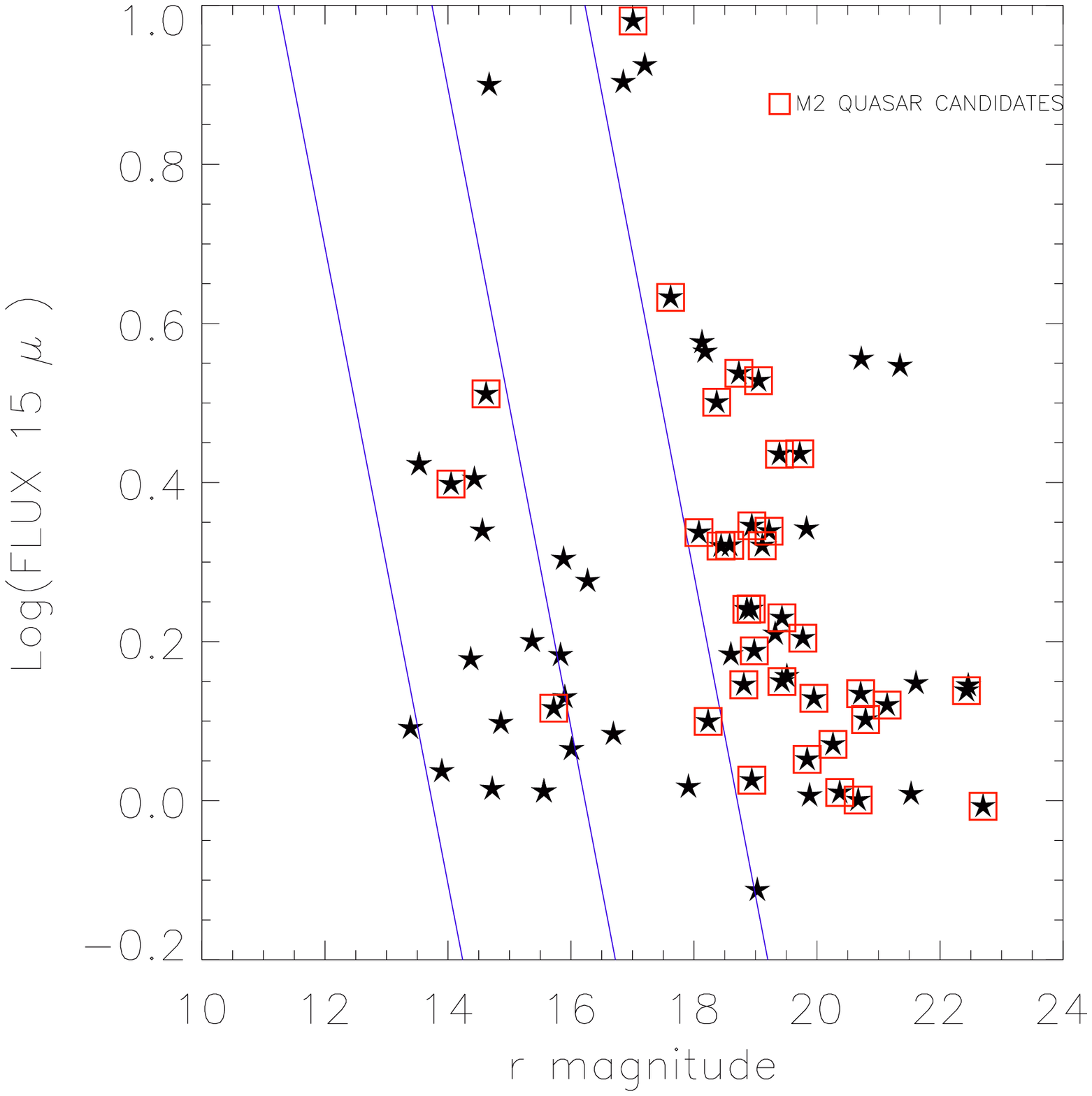,width=6cm,height=6cm}}
\caption{15 \mums flux (mJy) vs SDSS $r$-band magnitude for the point sources (stars)
within the parts of ELAIS {\it N1} and {\it N2} areas covered by SDSS. The red squares indicate
the confirmed quasars (left panel), {\it M1} quasar candidates (central panel) and {\it M2} quasar candidates 
(right panel). The blue lines represents constant 15 \mums to $r$-band flux, with values
0.1, 1., and 10.}
\label{log15rmag}
\end{figure*}

The 15 \mums information can be used in order to impose IR conditions in the selection
of quasar candidates.
Stars, galaxies, and AGN all have different optical to mid-IR slopes, with stars
typically having larger optical than mid-IR fluxes \citep{gonzalez04}. 
Furthermore, according to models of galaxies in the IR
\citep{rowan01}, quasar mid-IR fluxes are some 10 to 100 times 
larger than their optical ones. Taking this into account, one can impose additional constraints
on the selection criteria requiring a mid-IR to optical flux ratio of at least 10 for
quasar candidates (hereafter condition {\it C1}).
This condition allows the removal of three quasar candidates selected by $M2$
that have mid-IR to optical fluxes those of stars. Fig. \ref{log15rmag} illustrates the
positions of the different objects types and the regions where quasars and quasar candidates
selected by the two methods lie.

\begin{figure*}
\centerline{
\psfig{file=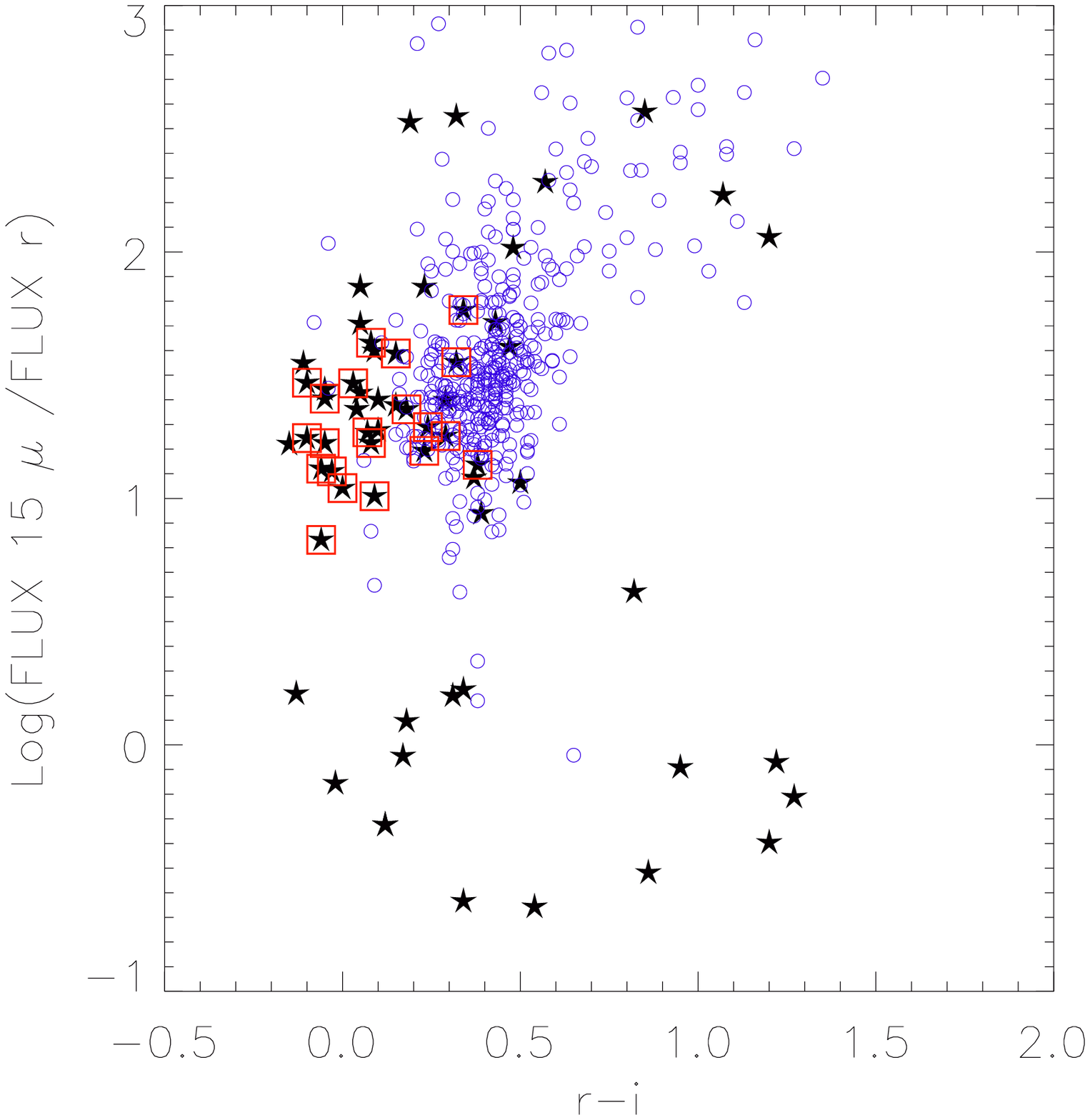,width=6cm,height=6cm}
\psfig{file=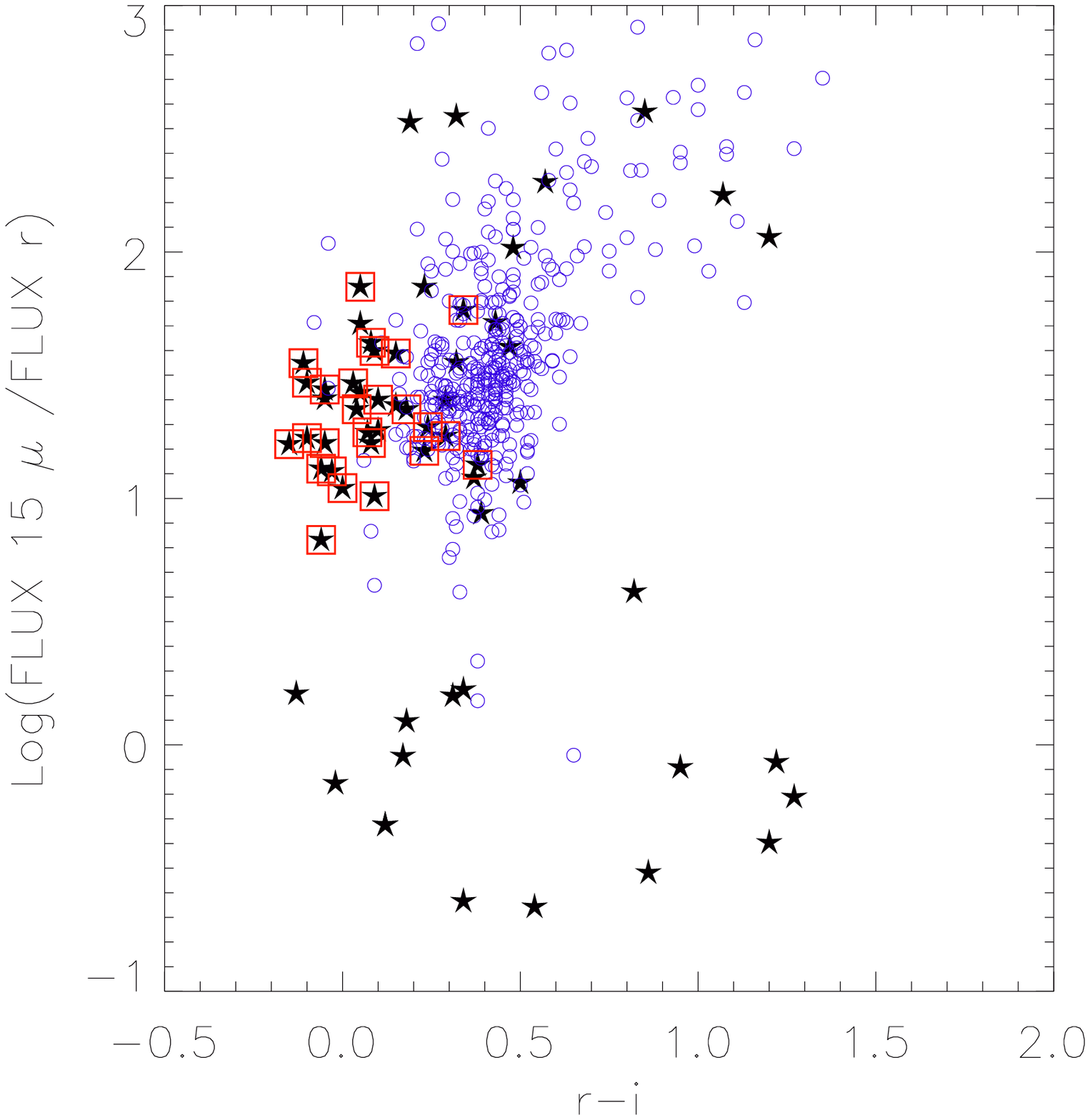,width=6cm,height=6cm}
\psfig{file=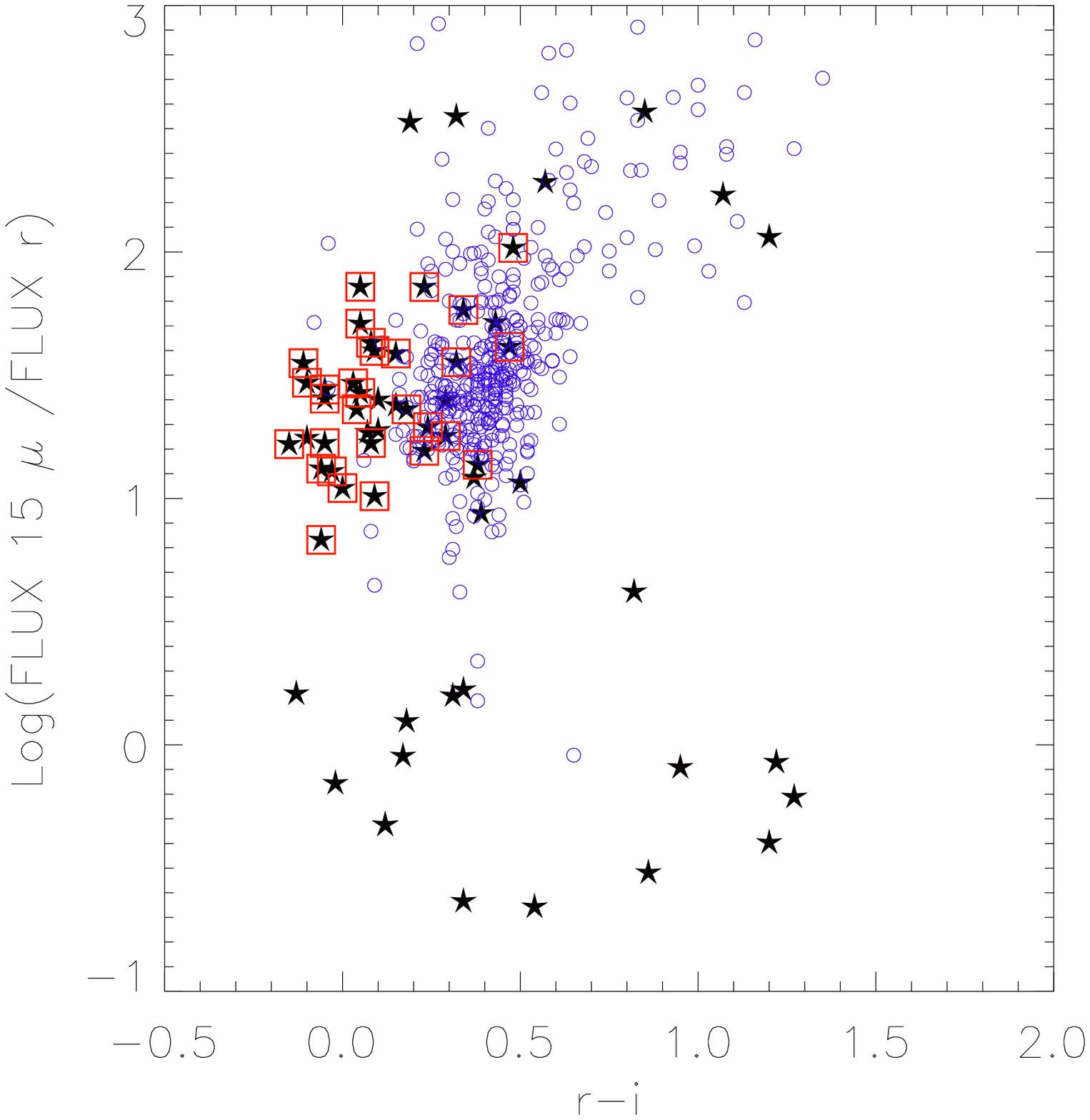,width=6cm,height=6cm}}
\caption{15 \mums to $r$-band flux vs SDSS $r-i$ for the point sources (stars)
and extended sources (open circles). Red squares indicate
spectroscopically confirmed quasars (left panel), {\it M1} quasar candidates (central panel) 
and {\it M2} quasar candidates (right panel).}
\label{iroptcolour}
\end{figure*}

In order to distinguish between quasars and galaxies, one can make a combined use 
of optical and optical/IR colours taking advantage of the fact that quasars up to
a redshift of $\sim 3$ are typically bluer than galaxies \citep{gonzalez04}.
Furthermore, 97.55\% of all spectroscopically confirmed quasars in the DR1 quasar catalogue
\citep{schneider04} have
$r-i < 0.52$ (hereafter {\it C2}). Fig. \ref{iroptcolour} shows the distribution of stars (lower mid-IR
to optical fluxes), galaxies and quasars (bluer than galaxies in general).
Marked in red are the spectroscopically confirmed quasars (left panel) and the candidates
selected by methods {\it M1} and {\it M2} (middle and right panels, respectively).
As can be seen, all spectroscopically confirmed quasars
form a clump bluewards of $r-i \sim 0.52$. In this particular case, {\it C2} does not improve
the results of any of the methods but will be used further on. 

If $N_c$ is the number of quasar candidates stemming from
the identification technique, $N_f$ the number of real quasars among
the candidates, and $N_e$ the number of expected (based on models) or known 
(based on complete observations) quasars,
{\it completeness} and {\it confirmation rate} can be
defined as ${N_f}$/${N_e}$ and ${N_f}$/${N_c}$, respectively \citep{hatzi00}.
Table \ref{tabcand} compares the two quantities yielded by the two methods.
Both methods give the same completeness but the colour-colour selection favours the 
confirmation rate.
Note, however, that the values given here for confirmation rate are
{\it lower limits}. A substantial number of
candidates (some 30\%) are fainter than the SDSS spectroscopic completeness limit
or lie outside the area covered by spectroscopy and therefore their nature is unknown.
For the same reason, the value of the completeness is also indicative as a complete
spectroscopic coverage would alter both $N_f$ and ${N_e}$.

\subsection{Quasar Candidates with WFS Photometry}   
\label{wfscand} 

\begin{figure*}
\centerline{
\psfig{file=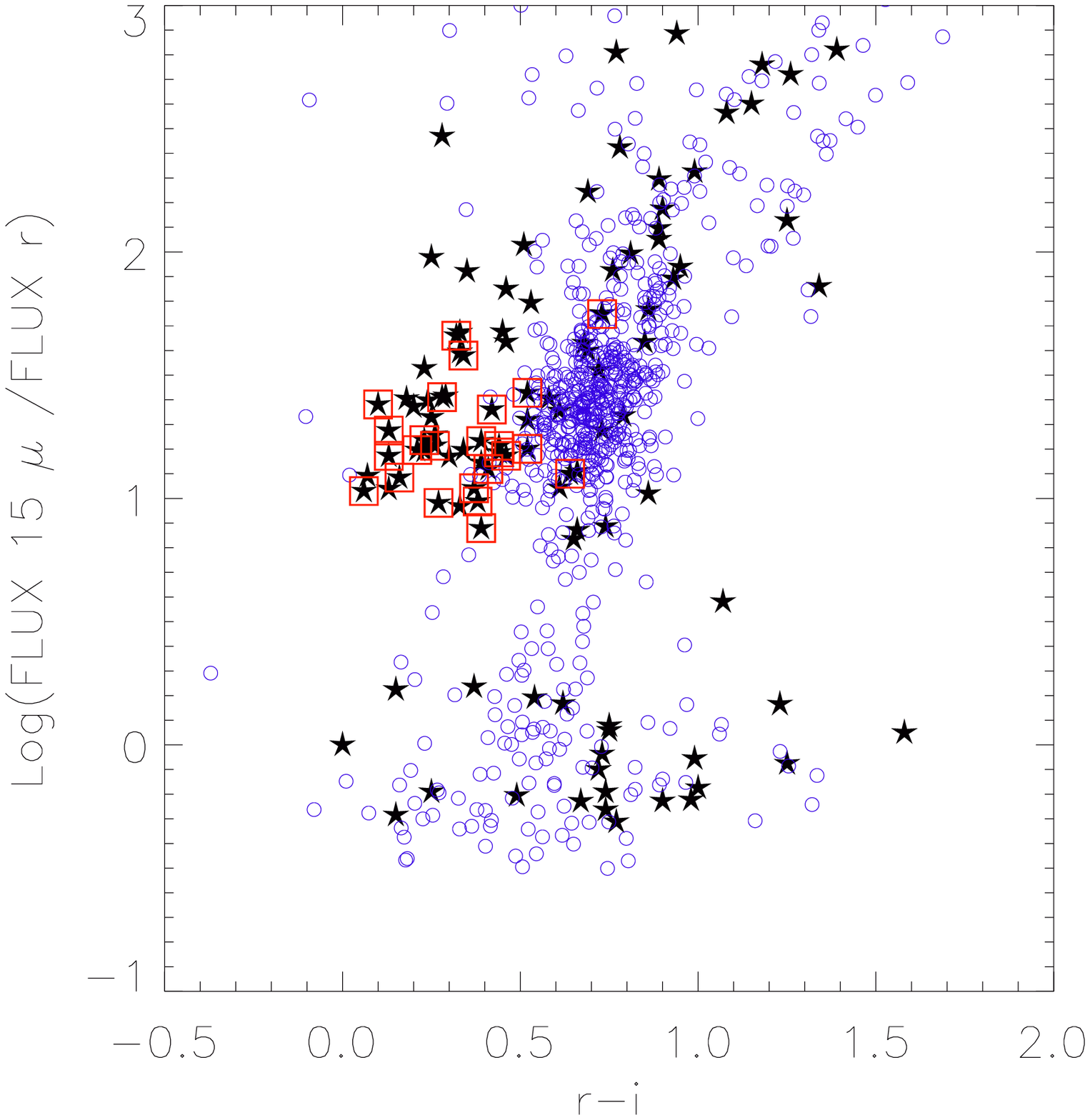,width=6cm,height=6cm}
\psfig{file=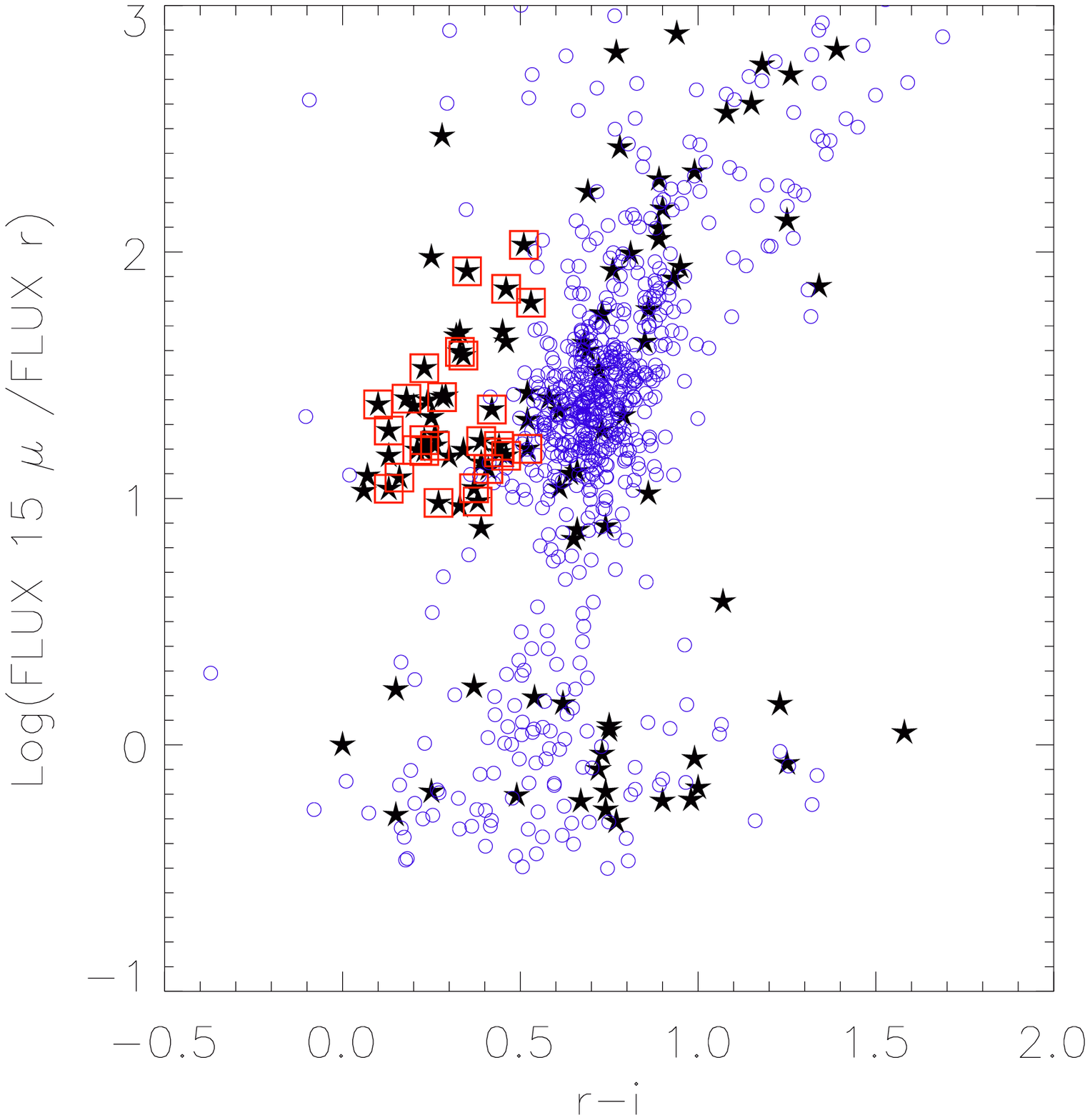,width=6cm,height=6cm}
\psfig{file=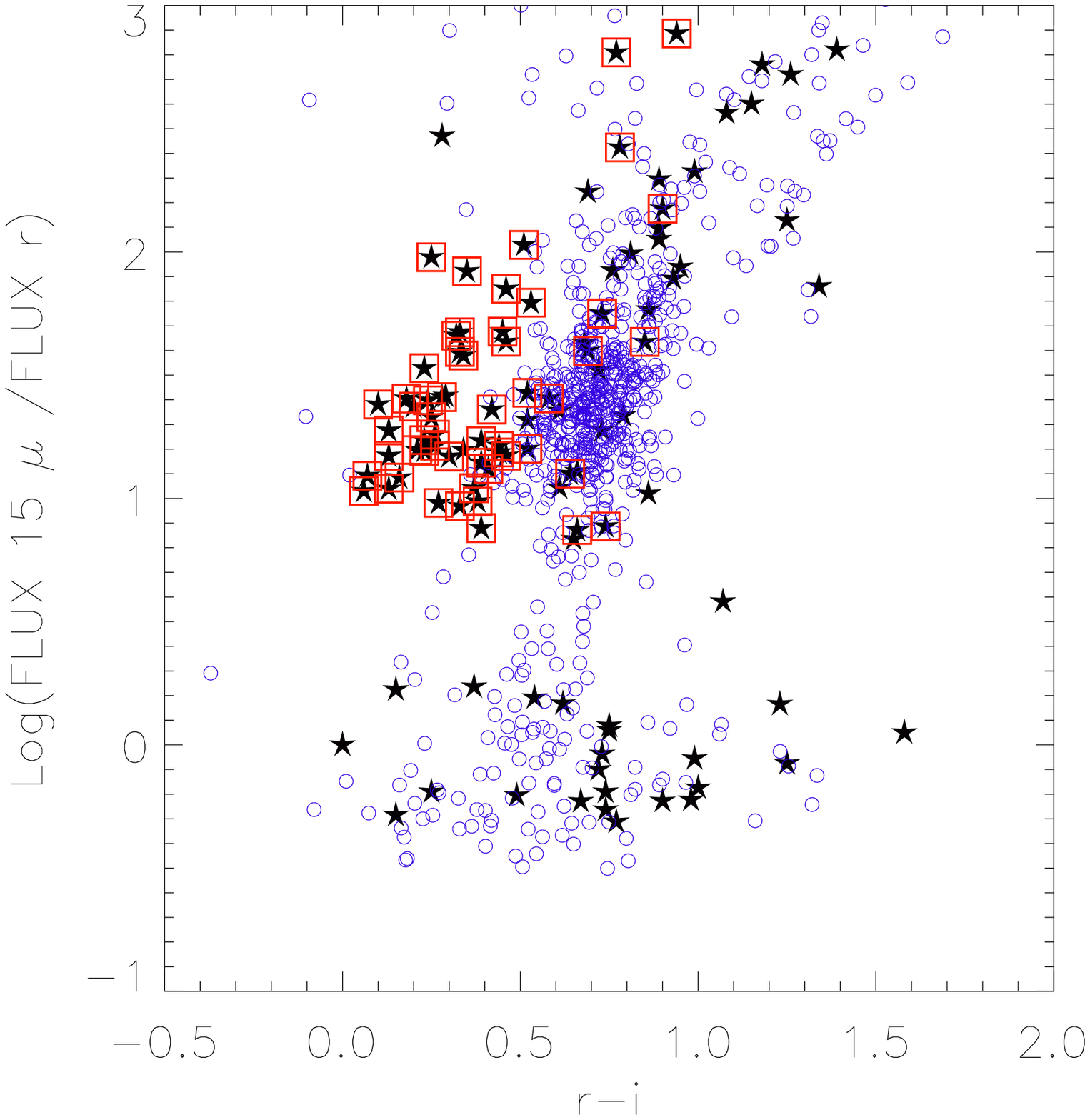,width=6cm,height=6cm}\qquad}
\caption{15 \mums to $r$-band flux vs WFS $r-i$ for the point sources (stars)
and extended sources (open circles). In red squares, the spectroscopically confirmed
quasars (left), {\it M1} quasar candidates (centre) and {\it M2} quasar candidates (right).}
\label{iroptcolourwfs}
\end{figure*}

From the objects morphologically identified as point sources (SExtractor CLASS\_STAR $\ge 0.9$)
with detections at 15 \mums 110 objects are part of the ELAIS Final Band-merged
Catalogue \citep{rowan03}. {\it M1} yields 27 quasar candidates with 18 spectroscopically confirmed, 
while {\it M2} finds 63 candidates with 25 spectroscopically confirmed (Table \ref{tabconfquasars}). 
The two candidate lists have 27 objects in common, 18 of which are
spectroscopically confirmed quasars. Taking into account condition {\it C1}, 
7 out of the 63 candidates proposed by {\it M2} can be safely discarded as stars.
Applying {\it C2} on the remaining 56 candidates, we discard another five as
more likely to be galaxies. Fig. \ref{iroptcolourwfs} shows the positions of
confirmed quasars and candidates, similar to Fig. \ref{iroptcolour}, but for the WFS photometry.

The results obtained using WFS photometry are summarised in Table \ref{tabcandwfs}. 
As already mentioned, the values of
completeness are underestimated due to lack of complete spectroscopic coverage. 

\begin{table}
\caption{Comparison of the completeness and confirmation rate yielded by {\it M1} and {\it M2} on
a sample of 110 WFS point sources with 15 \mums emission.
The second column shows the number of quasar candidates ($N_c$) and the number of 
spectroscopically confirmed quasars among them ($N_f$).}
\label{tabcandwfs}
\begin{tabular}{lccc}
\hline
\hline
Method & $N_c$ ($N_f$) & Completeness & Conf. Rate \\
\hline
\hline
{\it M1}     & 27 (18)  & 72\%         & 67\%       \\
{\it M2}     & 63 (25)  & 100\%        & 42\%       \\
\hline
{\it M2} \& {\it C1} & 56 (25) & 100\%     & 45\%       \\  
{\it M2} \& {\it C1} \& {\it C2} & 51 (25) & 100\%         & 49\%       \\
\hline
\hline
\end{tabular}
\end{table}

After thorough consideration we reach the conclusion that WFS data can be used 
in order to reliably obtain quasar candidates despite the issues raised by variability,
especially if one combines {\it M1} and {\it M2} with the constraints imposed by the objects' IR properties.

\subsection{Quasar Photometric Redshifts}
\label{zphot}         

\begin{figure*}
\centerline{
\psfig{file=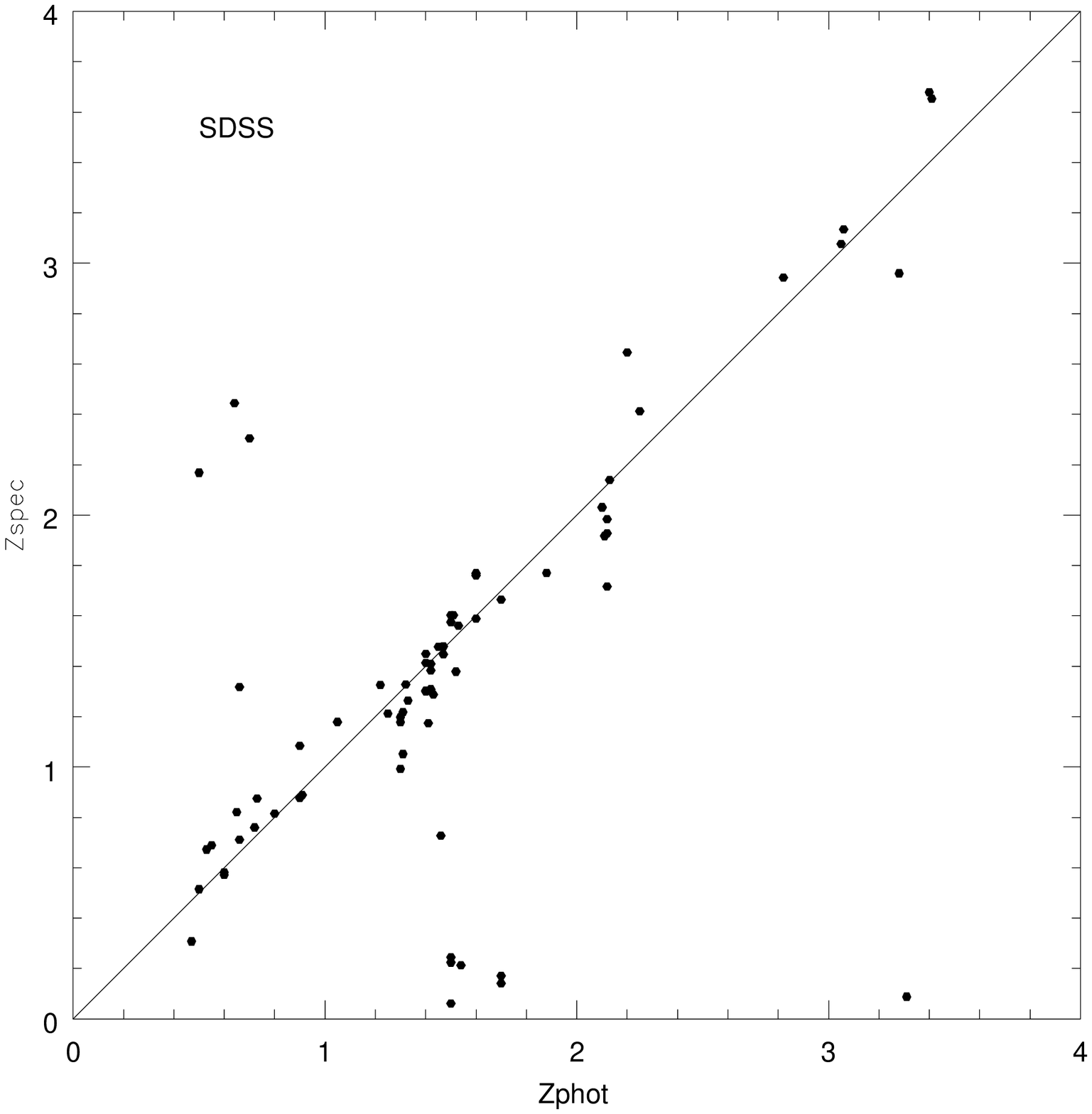,width=9cm,height=9cm}
\psfig{file=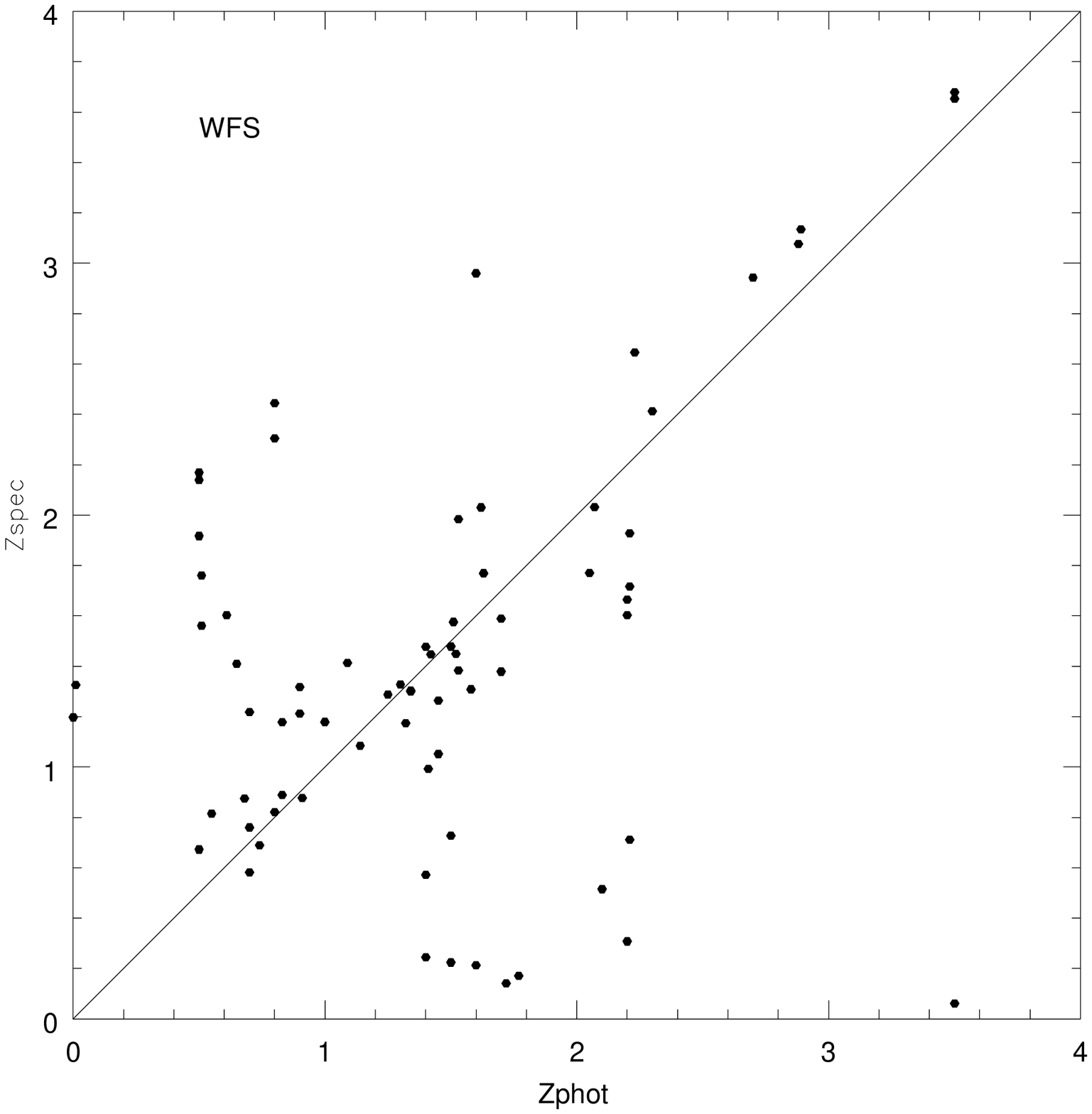,width=9cm,height=9cm}\qquad}
\caption{Photometric versus spectroscopic redshifts for the 73
SDSS quasars within the ELAIS {\it N1} and {\it N2} areas, using SDSS (left) and WFS (right)
photometry. The larger scatter in the right plot is most likely due to quasar variability,
see text for details.}
\label{figzphot}
\end{figure*}

For estimating the photometric redshifts we applied a standard template fitting procedure
using synthetic quasar spectra consisting of a power law continuum and
emission lines of fixed equivalent width values \citep{hatzi00}, 
on the sample of 73 spectroscopically confirmed quasars with both SDSS and WFS photometry available. 
The results are rather reliable when using SDSS
photometry (53 out of 73 $\Delta z \le$ 0.2, i.e. 73\% ), but get worse when using the
WFS photometry (29 out of 73 for $\Delta z \le$ 0.2, i.e. 30\% ). Note, however, that
all seven objects with spectroscopic redshifts lower than 0.3 have been assigned the wrong
photometric redshifts. These objects, however, are extended (OBJC\_TYPE = 3)
and, therefore, their magnitudes (psf
magnitudes for SDSS and core magnitudes for WFS) must be contaminated by the
light of the host galaxy. If we consider only the objects with
spectroscopic redshifts higher than 0.3, the numbers of good identifications become
80\% and 44\%, for SDSS and WFS, respectively. Furthermore, all 30 objects that were
assigned correct photometric redshifts using the WFS photometry have also correct
photometric redshifts when SDSS photometry is used. Fig. \ref{figzphot} 
illustrates the results obtained for SDSS (left panel) and WFS (right panel) photometry.

The fact that the bands for WFS photometry were taken in different periods of time 
(especially the $U$-band, which was taken more than two years apart in some cases) 
lead us to the conclusion that variability might be the basis of the
discrepancy problem we encountered, as described in Section \ref {variability}.

A case-by-case study of all objects that have correct photometric redshift estimates with
SDSS photometry and bad estimates ($\Delta z > 0.25$) with WFS photometry showed that they all
lie in the redshift range [0.3,2.0] and
their differences in magnitudes span a much larger range than the one expected due to
the filters' differences. One finally concludes that the wrong photometric redshift assignments
are most probably due to variability, as argued earlier.

\section{COMPARISON BETWEEN IR DETECTED AND NON-DETECTED QUASARS}
\label{comparison}   

\begin{figure*}
\centerline{
\psfig{file=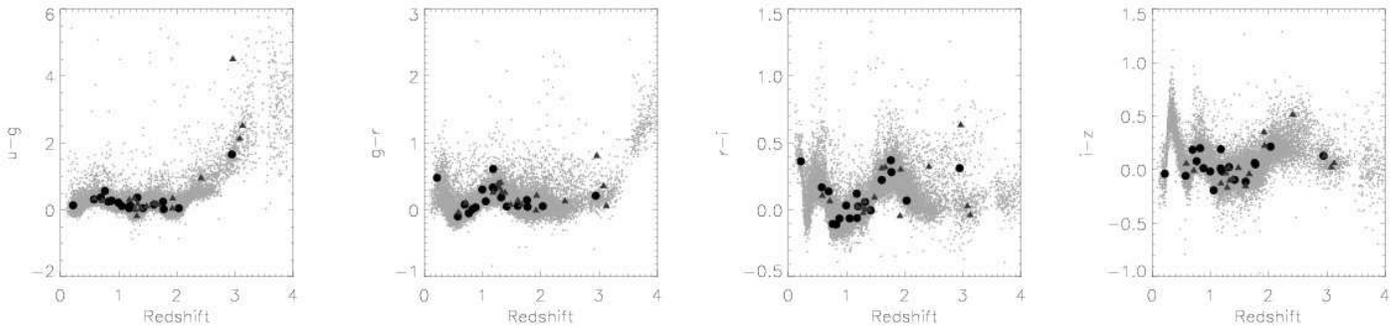,width=19cm}\qquad}
\caption{Colour - redshift diagrams for SDSS DR1 quasars (dots), 
non-ELAIS quasars (triangles) and ELAIS quasar (filled circles).}
\label{elaisnoelais1}
\end{figure*}

\begin{figure*}
\centerline{
\psfig{file=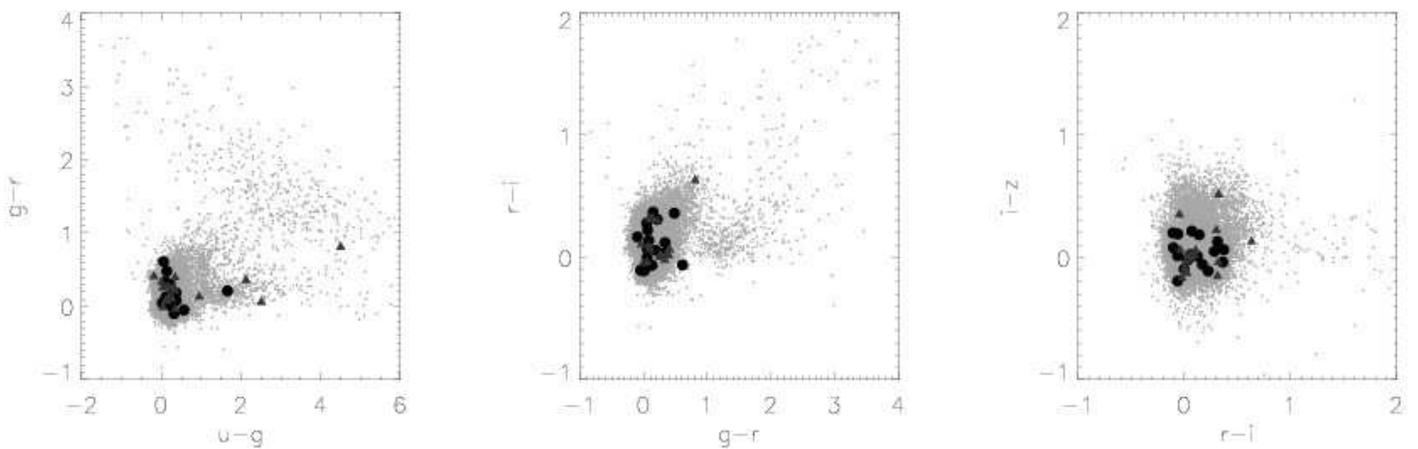,width=19cm}\qquad}
\caption{Colour - colour diagrams for SDSS DR1 quasars (dots), non-ELAIS 
quasars (triangles) and ELAIS quasar (filled circles).}
\label{elaisnoelais2}
\end{figure*}

36 spectroscopically confirmed quasars lie in the parts of the ELAIS {\it N1} and {\it N2} 
fields covered by the SDSS spectroscopic data release, with 16 of them detected at
15 \mum, as already mentioned. 
Considering the homogeneous way the SDSS quasar candidates are selected 
\citep{richards02} we can ask is if the two subsamples (16 IR and 20 non-IR emitters)
have the same properties or if the IR detected quasars are, in some way, different.
The colour-redshift (Fig.\ref{elaisnoelais1}) and colour-colour
(Fig.\ref{elaisnoelais2}) diagrams of the two subsamples do not show any differences.

\begin{figure*}
\centerline{
\psfig{file=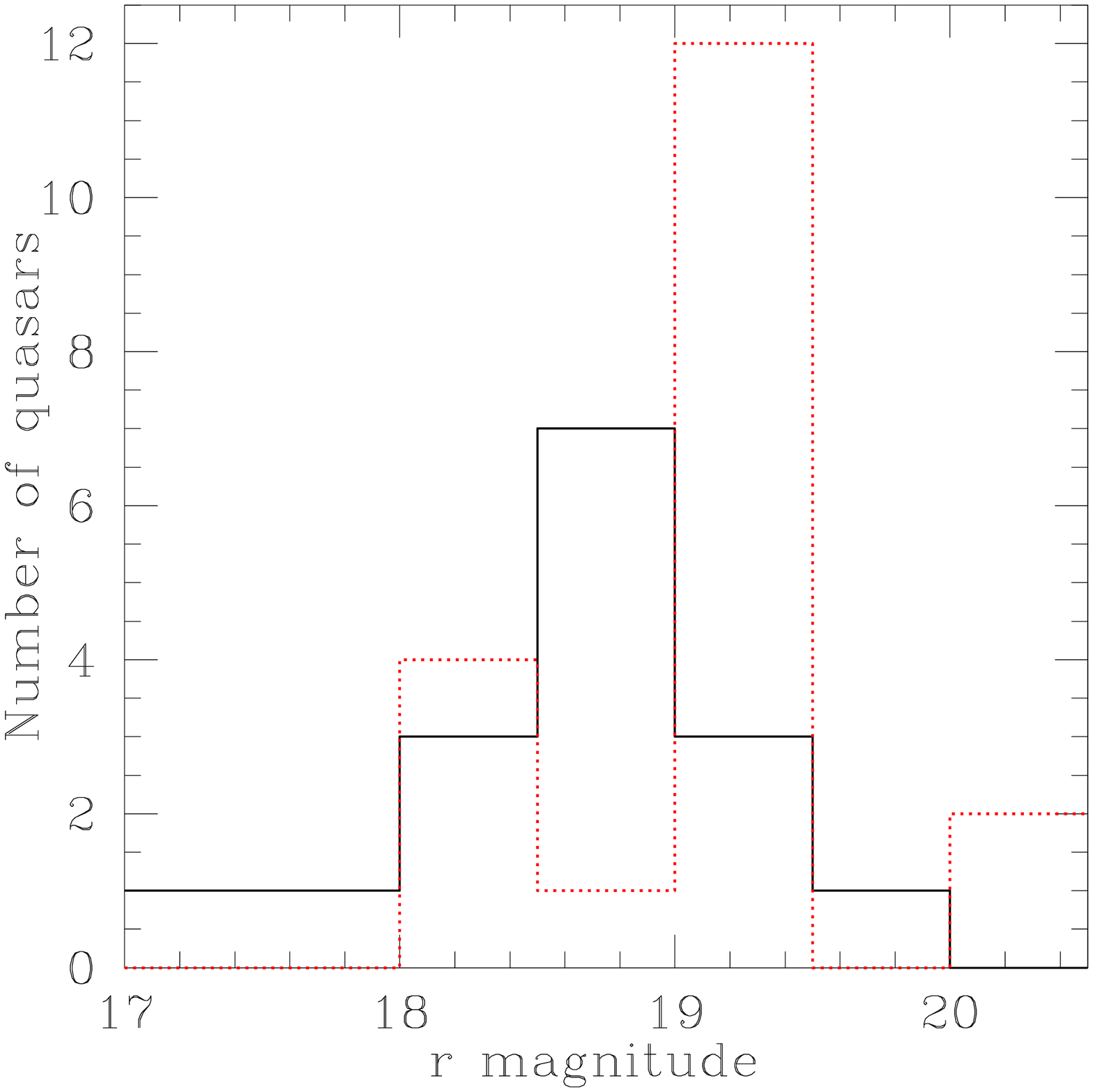,width=6cm,height=6cm}
\psfig{file=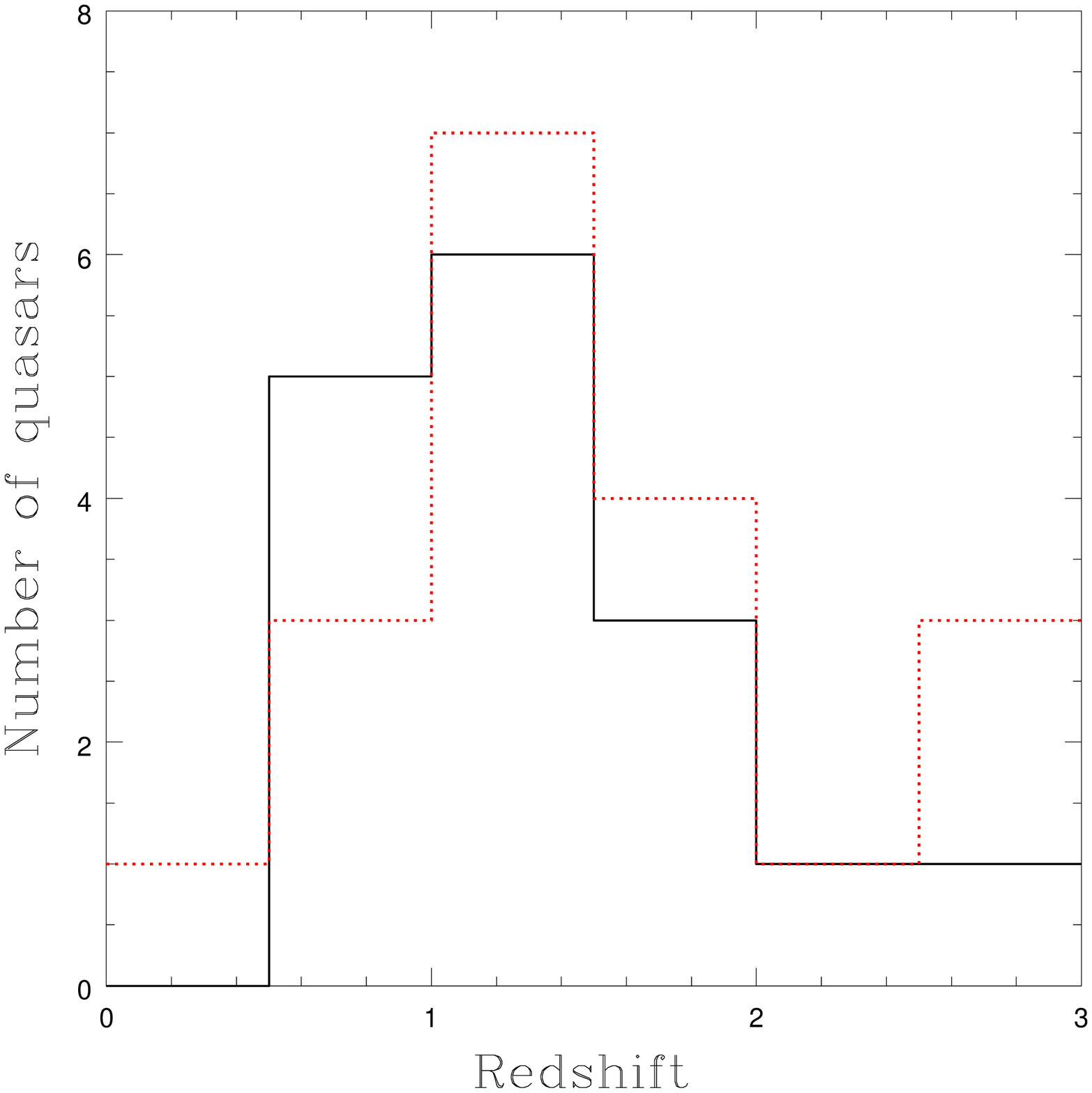,width=6cm,height=6cm}
\psfig{file=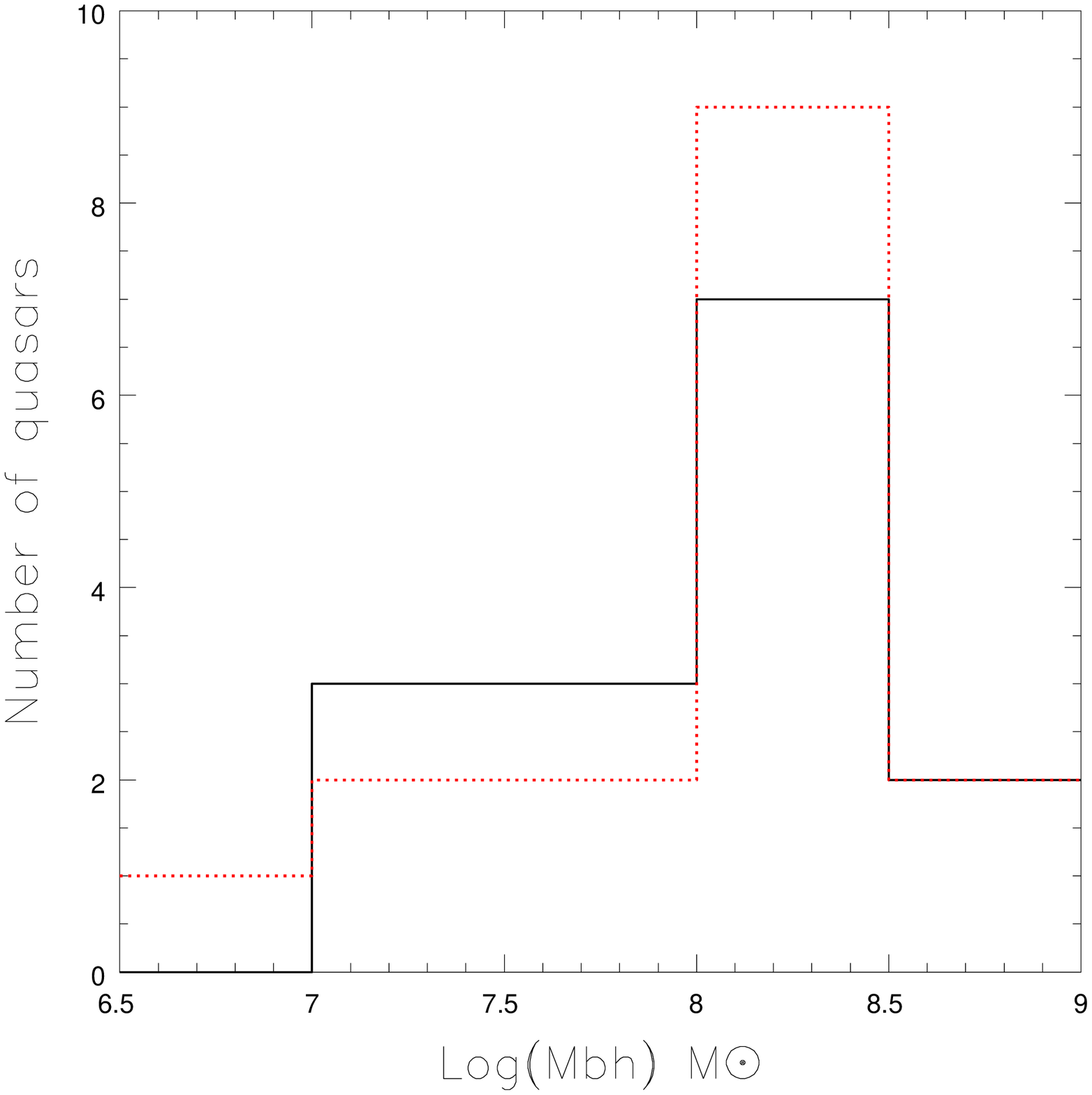,width=6cm,height=6cm}}
\caption{Histograms of $r$-band magnitude (left panel), redshift
(middle panel) and black hole mass (right panel) for the ELAIS (black solid line) 
and non-ELAIS (red dotted line) quasars.}
\label{fighistos}
\end{figure*}

A two-sided Kolmogorov-Smirnov test on the BH mass distributions of the two subsamples
(seen in the right panel of Fig. \ref{fighistos})
resulted in a larger than 90\% probability for them to come from the same population.
The same test gave a 35\% probability for the redshift distributions to be representative of
from the same population. The largest deviation is noticed in the $r$-band 
magnitude distributions (left panel of Fig. \ref{fighistos}).
The objects not detected by \isos at 15 \mums are on average half a magnitude
fainter in the $r$-band.  Fig. \ref{figflux} indicates a 
possible correlation between the 15 \mums fluxes of the SDSS quasars and
their optical ($r$-band) fluxes,
suggesting that the lack of 15 \mums counterparts could be due to their fainter
magnitudes. Deeper IR observation would, presumably,
provide counterparts for the remaining objects, since for a given covering fraction
and varying bolometric luminosity, brighter optical AGN should have a higher mid-IR
emission.

\subsection{Quasar Black Hole Masses}\label{mass}   

The principal assumption underlying the BH virial mass estimate
is that the dynamics of the Broad Line Region (BLR) are dominated by the 
gravity of the central supermassive black hole. Under this assumption an estimate 
of the central BH mass, $M_{BH}$, can be $M_{BH}\simeq R_{BLR}V^2/G$; where
$R_{BLR}$ is the radius of the BLR and $V$ is the velocity of the
line-emitting gas, traditionally estimated from the
FWHM of the $H\beta$ emission line (see \citealt{kaspi00}). For quasars with 
redshifts higher than typically $z \sim 0.8$, when $H\beta$ is no longer available,
\cite{mclure02} suggested the use of $Mg II$ as an estimator of the BH mass.
More precisely, the BH mass is computed as follows \citep{mclure04}:
\begin{equation}
\frac{M_{bh}}{M_{\odot}} = 4.7 \left(\frac{\lambda L_{5100}}{10^{37}W}\right)^{0.61} \left(\frac{FWHM(H_{\beta})}{km/sec}\right)^2
\label{eqHb}
\end{equation}
and
\begin{equation}
\frac{M_{bh}}{M_{\odot}} = 3.2 \left(\frac{\lambda L_{3000}}{10^{37}W}\right)^{0.62} \left(\frac{FWHM(MgII)}{km/sec}\right)^2 
\label{eqMg}
\end{equation}

For the sample of 36 SDSS quasars in {\it N1} and {\it N2}, therefore,
low redshift (up to $\sim 0.8$) mass estimates are based on the correlation 
between the $H\beta$ and the monochromatic luminosity at 5100 \AA \, (Eq. \ref{eqHb})
while for higher redshift ones the correlation between $Mg II$ and 3000 \AA \,
luminosity (Eq. \ref{eqMg}) is used. For objects with redshifts higher than 2, all estimators fall
outside the covered wavelength range and BH masses can no longer be computed. Only objects with SDSS
spectra were used for the BH mass computation as line and continuum of all SDSS
spectra are measured in a consistent way by the SDSS pipeline.

\begin{figure}
\centerline{
\psfig{file=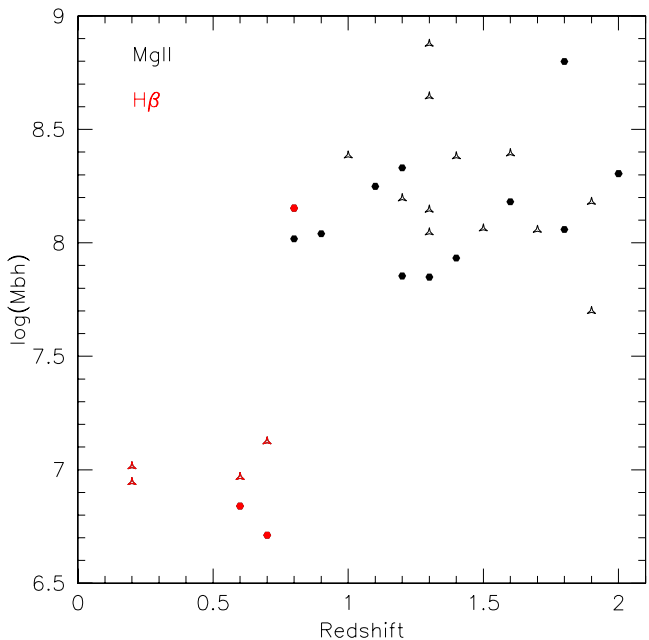,width=9cm,height=9cm}}
\caption{Black hole mass as a function of redshift for the ELAIS quasars
with SDSS spectra (filled circles) and for the SDSS quasars without
ELAIS counterparts (triangles).}
\label{figbhlum}
\end{figure}

Fig. \ref{figbhlum} shows the distribution of 
BH mass as a function of redshift. Red marks the objects for which $H\beta$
was used (with redshifts typically lower than $\sim 0.8$),
while black illustrates the ones for which $Mg II$ was considered. Filled circles denote the
objects with 15 \mums emission while triangles correspond to SDSS objects within the ELAIS
fields but without IR detections. BH masses show no differences between
the two subsamples and follow the distributions found for the entire SDSS DR1 quasar
catalogue by \cite{mclure04}. The right panel
of Fig. \ref{fighistos} shows the BH mass histogram for the two subsamples, clearly 
indicating that both of them belong to the same parent population.

\begin{figure*}
\centerline{
\psfig{file=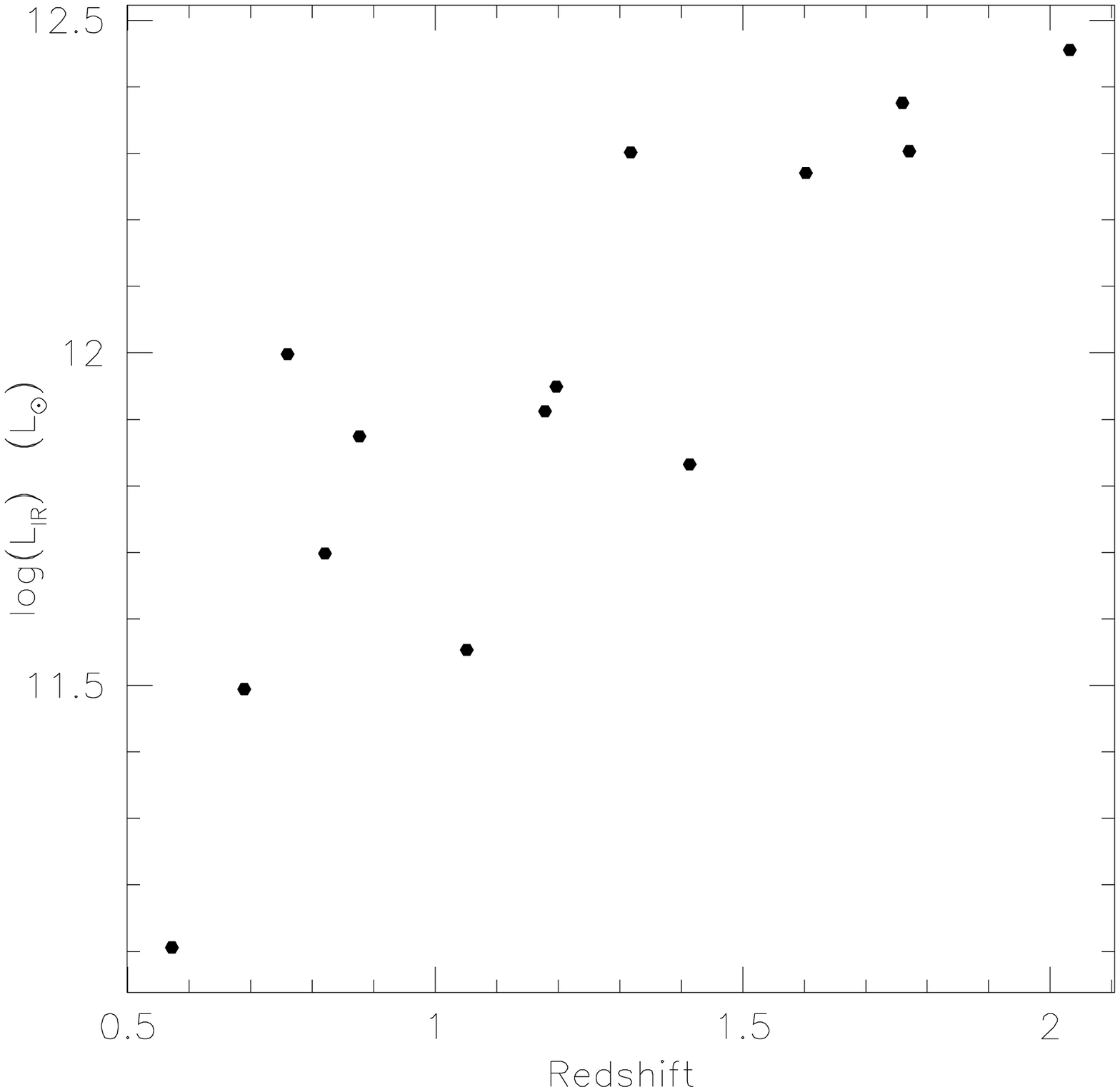,width=9cm,height=9cm}
\psfig{file=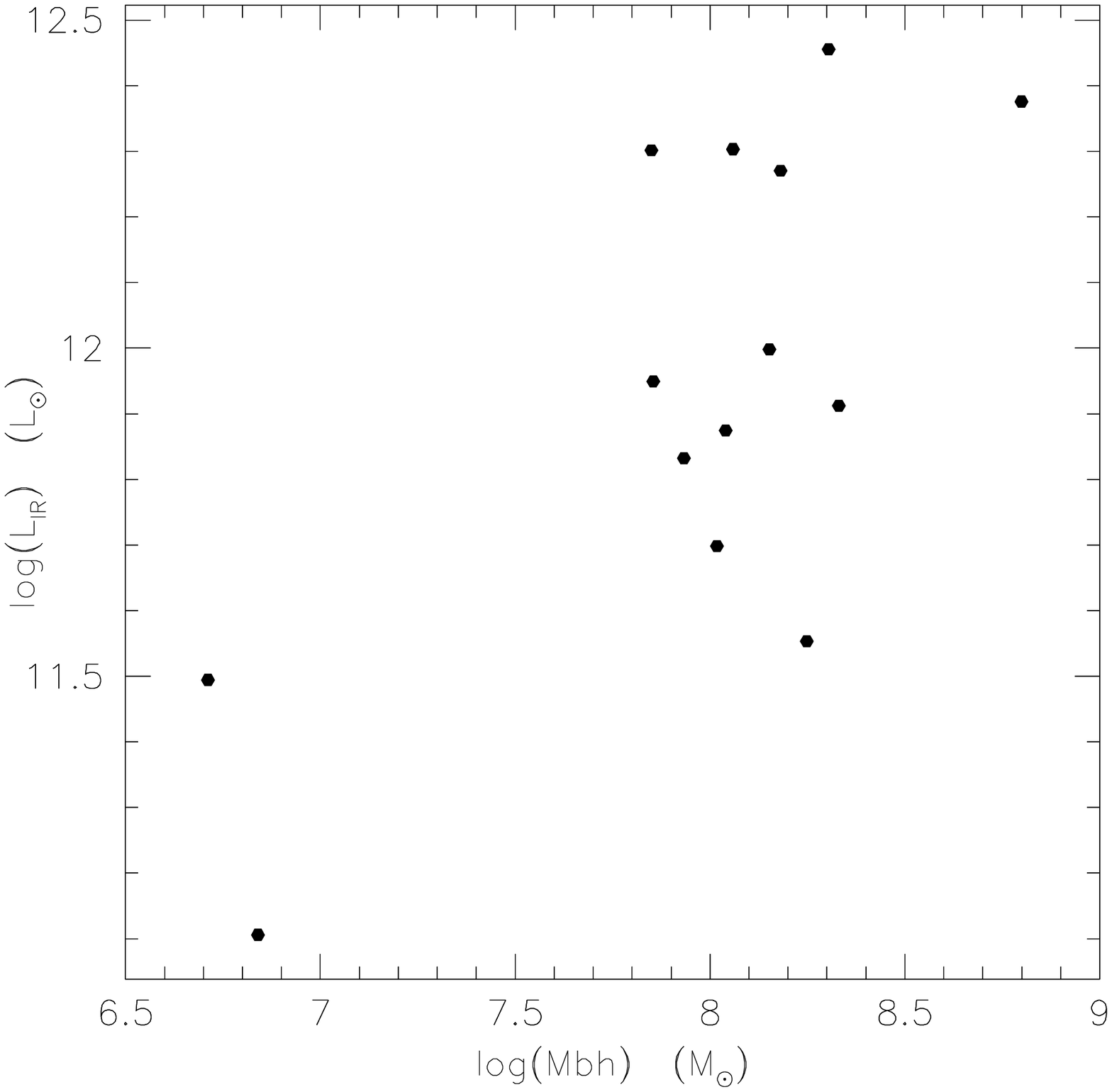,width=9cm,height=9cm}}
\caption{15 \mums IR luminosity of these objects as a function of redshift (left panel) and as
a function of the BH mass (right panel).}
\label{figIRlum}
\end{figure*}

For the objects for which the 15 \mums fluxes were available, their IR luminosity was
computed assuming an ($\Omega$, $\Lambda$) = (0.3, 0.7) cosmology. Fig. \ref{figIRlum}
shows the 15 \mums IR luminosity of these objects as a function of redshift (left panel) and as
a function of the BH mass (right panel). Clearly, high-mass BH tend to produce higher IR
luminosities. Furthermore, the IR luminosity versus redshift distribution compares nicely 
to the findings of \cite{pozzi03}, indicating that type I AGN in the redshift range [1,2]
have IR luminosities higher than 10$^{12} L_{\odot}$ (see Fig. 6 of \citealt{pozzi03}), 
proving thus that \isos only skimmed the brighter end of the luminosity function.

\section{DISCUSSION}   
\label{discuss} 
This paper discusses some properties of the ELAIS 15 \mums quasars and tries to establish
a robust method of quasar selection for future use within the SWIRE framework. The importance
of good quality ground-based data is stressed, both for the candidates selection and for
the photometric redshifts estimates. Colour-colour plots and template fitting are used for 
these purposes. Colour-colour plots give a higher confirmation rate, as expected since
they were trained on the same photometric system, while template fitting has the advantage of
being independent of the filters used.
The SDSS data have proved to be a reliable dataset, 
while the WFS data are somewhat less efficient due to the large timescales over
which the imaging has been carried out and the variability issues this introduced. In order
to correct for these effects, the IR and optical to IR properties of the objects have been
taken into account imposing additional constraints on the quasar candidate selection techniques.
The two subsamples of spectroscopically confirmed quasars detected and non-detected at 15 \mums
have been examined and they have shown no intrinsic physical differences.
Their non-detection is most probably due to their fainter magnitudes, probably correlated to
their higher redshifts.

This work can be seen as a validation of the tools and methods that will be used in the
framework of SWIRE or other similar IR surveys supported by ground-based optical data.
SWIRE will survey six high-latitude fields (including ELAIS {\it N1} and {\it N2}), 
totaling 50 square degrees in all seven Spitzer bands. One of the key scientific goals of 
SWIRE is to determine the evolution of quasars in the redshift range 0.5$ \le z \le $3 
\citep{lonsdale04}. 
As we suggested in Section \ref{comparison}, with deeper IR observations one could detect all the 
quasars in the discussed fields down to the optical magnitude limit. 
In particular, the SWIRE 5$\sigma$ photometric sensitivity
is 0.0037 mJy in the IRAC 8 \mums band and 0.15 mJy in the MIPS 24 \mums band, 
much deeper than the characteristic depth of $\sim 1$ mJy of ELAIS 15 \mums \citep{vaccari04}.
Last but not least, the multitude of the SWIRE bands will allow for a much better
galaxy-AGN separation in the mid-IR colour space, possibly making the optical morphological
preselection of quasar candidates obsolete.

Models (e.g. \citealt{granato94}; \citealt{nenkova02}) and observations
(e.g. \citealt{elvis94}) suggest that quasar IR spectra are more or less
flat (in $\nu L_{\nu}$) from $\sim 1$ \mums down to at least 25 \mum, even in the absence of starburst activity.
Therefore, the deeper and better quality SWIRE 8 \mums and 24 \mums observations will 
allow the detection of high numbers of quasars and the easy adaptation of the tools 
presented here. The stellar contamination
will probably be higher at 8 \mums than at 15 \mums but this problem will most likely be solved
due to the multitude of IR bands, that will allow an easier separation of the galactic 
and extragalactic populations.

As a last remark, we would like to stress that a robust candidate selection 
technique and subsequent photometric redshift estimates such as the ones presented here
will be increasingly required by all future large area surveys, as spectroscopic
coverage will never reach the same completeness, neither in area nor in depth.

\vspace{0.75cm} \par\noindent
{\bf ACKNOWLEDGMENTS} \par

\noindent This paper is based on observations with \iso, an ESA project,
with instruments funded by ESA Member States and
with participation of ISAS and NASA. This work made use of
data products provided by the CASU INT Wide Field Survey and the Sloan
Digital Sky Survey. The INT and WHT telescopes are operated on the
island of La Palma by the Isaac Newton Group in the Spanish Observatorio
del Roque de los Muchachos of the Instituto de Astrofisica de Canarias.
The SDSS Web site is http://www.sdss.org/.
This work was supported in part by the Spanish Ministerio de
Ciencia y Tecnologia (Grants Nr. PB1998-0409-C02-01 and ESP2002-03716) 
and by the EC network "POE'' (Grant Nr. HPRN-CT-2000-00138).

\end{document}